\documentclass[10pt]{article}

\usepackage{graphicx}
\usepackage[margin=2cm]{geometry}
\usepackage{hyperref}
\usepackage{natbib}
\usepackage{amsmath,amssymb}
\usepackage{multirow}

\DeclareMathOperator{\E}{\mathbb{E}}
\DeclareMathOperator{\Proba}{\mathbb{P}}
\newcommand{\Eb}[1]{\ensuremath{\E\!\left[#1\right]}}
\newcommand{\Pb}[1]{\ensuremath{\Proba\!\left[#1\right]}}
\newcommand{\norm}[1]{\left\lVert #1 \right\rVert}

\title{Introducing endogenous transport provision in a LUTI model to explore polycentric governance systems}

\author{Juste Raimbault$^{1,2,3,\ast}$ and Florent Le N{\'e}chet$^{4}$\medskip\\
$^{1}$ Center for Advanced Spatial Analysis, University College London\\
$^{2}$ UPS CNRS 3611 ISC-PIF\\
$^{3}$ UMR CNRS 8504 G{\'e}ographie-cit{\'e}s\\
$^{4}$ LVMT, Universit{\'e} Gustave Eiffel\medskip\\
$^{\ast}$ \texttt{juste.raimbault@polytechnique.edu}
}
\date{}

\begin{document}

\maketitle

\begin{abstract}
Models focusing on interactions between land-use and transportation mostly assume an exogenous provision of transportation infrastructures. We investigate here co-evolutionary processes between land-use and transportation, at the scale of Mega-City Regions, by introducing a toy model of corresponding processes. In particular, our model is specifically tailored to include governance processes ruling the growth of transportation infrastructure. We show through stylised numerical simulations the potentialities of our model to reproduce a variety of dynamics when co-evolution is taken into account. We then apply the model to a case study, by calibrating it for the Pearl River Delta Mega-city Region (China, 1990-–2010). To go beyond this first modelling step, we elaborate on the challenges to overcome to go further towards more complex models integrating co-evolution between transportation networks and territories in urban systems.\\\medskip

\end{abstract}

%%%%%%%%%%%%%%%%%%%%%%
\section{Introduction}
%%%%%%%%%%%%%%%%%%%%%%

The emergence of Mega-City regions (MCR), which can be defined as a polycentric network of highly connected cities \citep{hall2006polycentric}, is since the past decades under the spotlight of various research domains \citep{gottmann1964megalopolis,xu2008planning, shao2009ground, pagliara2012megacities,taubenbock2014new}.  MCR goes beyond the traditional scale of cities. As stated by \cite{soja2000postmetropolis}: ``\textit{the urban, the metropolitan, and the subnational-regional scales seem to be blending together in many parts of the world}''. There is increasing need to understand the main processes occurring at urban and interurban scales, affecting the emergence and the functioning of such multi-scalar geographical objects \citep{rozenblat2020extending}. In this article, we explore the conditions of emergence of an integrated governance for polycentric MCR, using a modelling approach tackling the co-evolution between transport and territory~\citep{levinson2011coevolution,raimbault:tel-01857741,mimeur2020analyse}. This reflects with empirical literature studying integrated governance at the metropolitan scale \citep{heeg2003metropolitan,lenechet2017peupler}, or even at wider regional or MCR scale \citep{neuman2009futures,xu2010mega}.

\subsection{Modelling transport / territory coevolution through governance}

Transport and cities have complex and intricate trajectories over time. Path-dependency is central in such systems, in the sense that some planning decisions will influence during decades the future choices of all agents involved. Hence, transportation networks built over long time carry the mark of scales at which the decision were taken: the French railway network is a good illustration, as the “Etoile de Legrand”, planned in 1842 at national level led to a highly centralised network around Paris \citep{ribeill1985aspects}, although pre-existing postal roads networks were not centralised \citep{verdier2007extension}. This decision had consequences on network topology but also on the trajectory of the system of cities over the long term~\citep{bretagnolle2003vitesse}, sealing a strong position for Paris at national level. Therefore, the state of a network can be seen as a signature of past decisions.

Similarly, the transportation network between cities within a MCR both witnesses the scales at which transport provision was decided, and influences in return the future dynamic of the MCR. In this article, we adopt a modelling approach aiming at simulating the dynamic of MCR over the long term, with a special focus on the interaction between the location system and the transportation system~\citep{lowry1964model,bonnafous1996systeme}. Before presenting our research question, we wish to emphasise the diversity of modelling approaches that are used to tackle these interactions.

Mainly used in operational contexts, LUTI models were developed in the 1970's to extend the classical ``four-stage'' model of transportation, and include endogenous evolution of population and jobs distribution over space. This family of models \citep{ waddell2002urbansim,wegener2004overview,coppola2013luti,acheampong2015land,lopes2019luti} are used to assess the territorial effects of new links provision in a transportation network, for medium temporal scales (circa 20 years). However such models are not built to predict the evolution of transportation networks. This research area is not well covered by the literature with few exceptions such as \cite{russo2012unifying} which include macro-economic models of infrastructure provision but does not endogenise explicitly the spatial growth of the transportation network.

Models of co-evolution between land-use and transportation networks that aim at addressing those limitations are an emerging field, at the crossroad of different disciplines, such as transportation economics, physics, geocomputation. It has been suggested by~\cite{raimbault2017models} that disciplines are strongly fragmented when coming to study this transdisciplinary subject. This may be due to different research questions and applications, various spatial and temporal scales at stake and different objects of study. Geographers have produced some empirical studies~\citep{bretagnolle:tel-00459720} and models focusing on longer time scales that are fitted for system of cities scale: for instance the SimpopNet model~\citep{schmitt2014modelisation} aims at endogenising transportation network differentiation and hierarchisation, driven by the interactions between cities. Rather than obtaining realistic models, the focus shifts to the understanding of the respective roles of stylised processes such as endogenous growth, gravity flows within the network, or network evolution. A similar approach is adopted by physicists working on simple models of co-evolution, with a further focus on isolating minimalist models for example based on local geometrical optimisation~\citep{courtat2011mathematics}. Economists \citep{xie2009modeling} have also widely studied this question, their models putting the emphasis on the economic accuracy of network development processes (investment models, role of basic economic agents). Such a diversity of viewpoints highlights the difficulty of modelling interactions between transportation network and territories in an endogenous way.

One way to tackle the transportation/territory co-evolution processes is by including an explicit mechanism of transportation provision, that we call ``governance process''. Several approaches of co-evolution models including governance processes have been suggested in the literature. Hence, decisions to build transportation infrastructures are taken by stakeholders operating at different scales, from municipalities to states. Depending on negotiations between various actors, different decisions will create qualitatively different networks. \cite{li2016integrated} couple a network investment model with a traffic and location model, and show that the obtained steady state configurations outperform an operational research approach to network design in terms of overall accessibility. The example of Twin Cities, bi-centric with an intermediate airport, is used as an application case for a network growth model with explicit investment in \cite{zhang2016model}. Also regarding network growth, \cite{jacobs2016transport} describe a simulation model in which alternatives between plausible investments by different investors are assessed using a discrete choice model which utility function takes into account returns on investment but also variables to optimise such as accessibility. It is applied to the growth of the Dutch railway in the 19th century, and reproduces quite accurately the historical network. \cite{yusufzyanova2011forecasting} describe a model of network growth with investments, with which a full state regulation can be compared with a situation where investments are managed by local agencies.

We believe this emerging area of research to be complementary with the operational LUTI models, and that efforts to bridge gaps between detailed description of medium term effect of transportation infrastructure provision and long term co-evolution between transport and territory have potential to be used in the context of complex socio-technical transition of territories \citep{murphy2015human}. In this paper, we suggest a modelling framework that goes towards this direction, and we believe this framework to be useful to understand the complex trajectories of MCR.

\subsection{Transportation provision and emergence of Mega-City Regions}

Transportation networks play an important role in the emergence of MCR: such level of polycentricity and interlacing of geographical scales can be reached only with an important heterogeneity of travel speeds, for instance through High Speed Rail (HSR) networks between cities. European examples about the progressive interlacing of formerly independent job markets can be found in Italy after the creation of the HSR network between Rome and Naples \citep{cascetta2011analysis}, or in Northern Europe where the {\"O}resund bridge between Denmark and Sweden played a major role in fostering the emergence of an international Copenhagen-Malmo job market \citep{matthiessen2005oresund}. In both cases it is important to keep in mind that within a MCR the urban scale remain relevant to analyse the action of individuals: \cite{olesen2017region} underlines that the existence of a polycentric {\"O}resund region does not affect the persistence of Copenhagen and Malmo metropolitan regions as consistent territorial entities. MCR are not just wider metropolitan regions, they intrinsically imply multi-scalar dialog and coordination between stakeholders. Due to their construction and maintenance costs, transport infrastructures are increasingly financed with a participation of territorial stakeholders at all scales, from local to national and even international scale (see Trans-European Transport Network). Thus, the choice to build a new transport infrastructure can be described as a negotiation between stakeholders operating at various scales. Transportation systems are particularly sensitive regarding this question of governance structure within MCR \citep{evers2013explaining}. \cite{pemberton2000institutional} illustrates for example this issue empirically for a conurbation in North-East England, and concludes that a uniform policy at the country level fails to produce integrated transportation systems. Overall, this overview shows the diversity of factors influencing the scale at which decision are taken at MCR scale.

On the modelling side, taking into account such processes, \cite{Xie2011} introduce a theoretical economic model of infrastructure investment. Two levels of governance, local and centralised are considered in the model. For the provision of new infrastructure that has to be split between two contiguous districts (space being one-dimensional), a “game” in the sense of game theory between planning agents determines both the level of decision and the attribution of the stock proportion to each district. Governments either seek to maximise the aggregated utility (Pigovian government), or include explicit political strategies to satisfy a median voter. Numerical exploration of the model show that these processes are equivalent to compromises between cost and benefits, and that the level of governance depends on the state of the network. \cite{xie2011governance} develop a simpler version of this model on the governance side but coupled with a more realistic component for transportation. Their model integrates on a synthetic growing network a traffic model with a pricing model and an investment model. They show that under the assumption of centralisation, an equilibrium between demand and network performance can be reached, but that investments are not efficient on the long run, with a higher loss for decentralised investments. \cite{adelt2018simulation} have also explored the context in which coordination appears useful between various stakeholders in a urban context.

By focusing on transport/territory co-evolution processes occurring within MCR at middle and long term scale (typically 50 years) through a modelling approach, our research question is as follows: in a context of network growth, what are the spatial conditions for the emergence of an integrated transport provision governance? Which factors are critical to explain this potential emergence, for example the distance between cities or the preexisting hierarchy of the system of cities? By interrogating the relevance of the MCR scale, we try to further explore ``mega-city regions [as] an emerging scale of [...] spatial regulation'' \citep{xu2010mega}. Our main contribution in this regard is the introduction of a novel simulation model for the co-evolution of transport networks and territories at the MCR scale, integrating governance processes as network evolution rules. This model, named LUTECIA, extends the LUTI framework on longer time scales. We apply it to the case of Pearl River Delta in China, to explore its potential to reconstruct retrospectively the possible power relationship between planning agents of this highly polycentric urban region.

The rest of this paper is organised as follows: in the next section the formalisation of the model is detailed. We then develop numerical experiments to validate the model and extract stylised facts from its behavior. The model is finally applied to the case of Pearl River Delta, China, one of the most densely populated and dynamic MCR in the world. We finally discuss the implications of our findings about the emergence of integrated governance at MCR level, on one hand, and the modelling framework suited to such explorations, on the other hand.

%%%%%%%%%%%%%%%%%%%%%%%%%%%
\section{Description of the LUTECIA Model}
%%%%%%%%%%%%%%%%%%%%%%%%%%%

In this paper, we focus on the modelling of interactions between territories and transportation networks, using a ``toy model'' named LUTECIA. On the one hand, alike LUTI models, we try to model how new transportation links shape the dynamics of cities (e.g. distribution of population and employments through space). On the other hand, we aim at taking into account the other part of the transportation/territory loop by including the decision-making process for a new transportation network link. We do so while combining two scales of governance: the urban level and the MCR level.

This model extends in particular the model introduced by~\cite{le2010approche,lenechet:halshs-00674059}, which was then developed by~\cite{nechet2019modelling} and is a co-evolution model between transportation and land-use. It extends LUTI models, in the sense that infrastructure provision decisions are taken by explicit agents in the model. This model is positioned in a logic close to \cite{Xie2011} for the role of the governance structure, and close to \cite{xie2011governance} for the degree of realism of how space is included in the model. By adding a multi-scalar layer through governance processes occurring at various scales, we wish to test whether this helps to reproduce the complex multi-scalar networks observed in MCR around the globe \citep{lenechet2017peupler}.

\subsection{Model structure}

We now describe the LUTECIA model. The model couples a module for land-use evolution with a module for transportation network growth. The different sub-models, detailed in the following section, include in particular a governance module that accounts for network evolution. It is based on iterative increases of accessibility levels, a widely used concept in the literature \citep{farrington2007new} that we found to be practical to link the spatial organisation of cities with the geography of transportation networks. Compared to operational models used in the most recent LUTI literature \citep{acheampong2015land, lopes2019luti}, it is not data-driven but rather stylised: it does not take into account detailed activity programs, nor advanced measures of accessibility. Commuting is the only travel purpose that we included in the model, and population is not segmented based on social characteristics. We furthermore made the assumption that beside the provision of transportation infrastructures, which are in practice mostly funded by the public sector given the difficulty to make profits, the rest of the city has a fluid evolution. This is in the sense that micro-economic agents would adjust rationally to the new configuration of transportation network, including the production of dwellings and corporate real estate. These assumptions are driven by the structuring role of transportation infrastructures in urban development, at least during the early phases of network maturation - see for example \cite{hou2011transport} showing the attenuation over time of the effect of accessibility on productivity. These choices allow us to integrate medium term processes from LUTI models (urban sprawl, saturation of central networks) with long term co-evolution processes, and ultimately to tackle our research question, the conditions for the emergence of an integrated governance at the MCR scale.

In its more general structure, the LUTECIA model is composed by five sub-models:
\begin{itemize}
\item LU - Land Use module: it proceeds to the relocation of actives and employments given current conditions of accessibility.
\item T - Transport module: it computes the transportation conditions such as flows and congestion in the urban region, and finally accessibility.
\item EC - Evaluation of Cooperation module: it manages the agent that will proceed to the provision a new infrastructure, after having assessed their willingness to cooperate.
\item I - Infrastructure provision module: it determines the location of the new transportation infrastructure, based on a criteria of accessibility maximisation.
\item A - Agglomeration economies module: it evaluates the productivity of firms, depending on the accessibility to employments.
\end{itemize}

\begin{figure}
\centering
	\includegraphics[width=\linewidth]{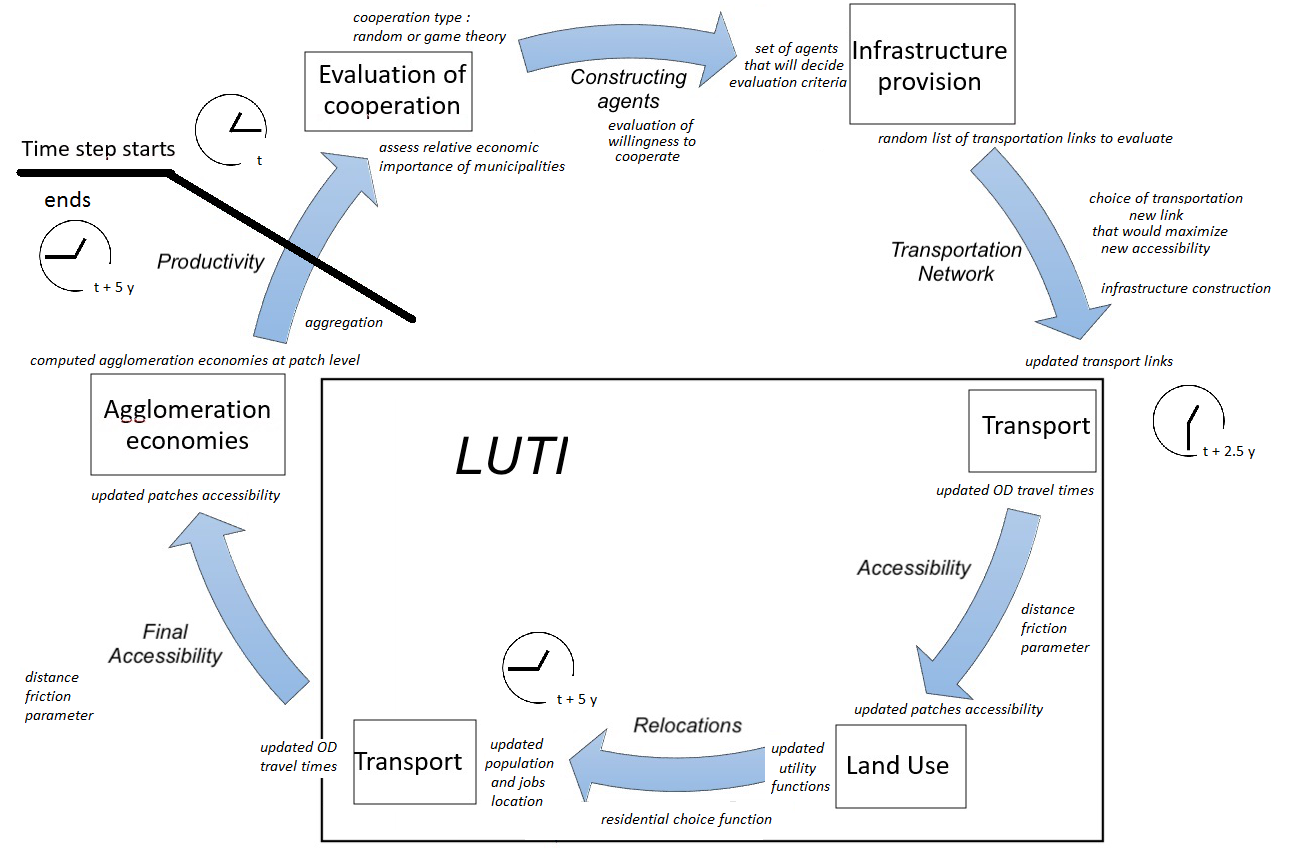}
	\caption{\textbf{Overview of LUTECIA sub-models.} We highlight the part corresponding to the traditional LUTI framework. We give crucial output of sub-models. Note that the transport module is called twice in this scheme, to compute new congestion patterns and new accessibility given land-use relocations, which have then a role in the evolution of infrastructure.\label{fig:submodules}}
\end{figure}

The articulation between the different sub-models via main inputs and outputs is shown in Fig.~\ref{fig:submodules}.

%%%%%%%%%%%%%%
%%%%%%%%%%%%%%

Different time scales are included in the model: a short scale, corresponding to daily mobility that yields flows in the transportation network, and to firms productivity (modules T and A); an intermediate to longer time scale for residential and firms dynamics (module LU); as well as for the evolution of the network (modules EC and I). We typically assume a time step to be of the order of magnitude of 5 years. To simplify the description of our model, we assume that the first medium time scale (provision of new transportation link) takes around 2/3 years, including studies, coordinated decision and completion of the network. This is indeed a very optimistic scheme, even for a motorway network, but we argue this is an acceptable framework in our modelling context. First, we do not include the existence of a large scale and long-term “master plan”. We assume a medium term rationality where infrastructure are built step by step. Also, we aimed via this choice to avoid a further complexity: indeed in reality there is always a time lag between the observation of a phenomenon and the decision making; there is also a time-lag between the decision and the actual provision of the infrastructure. Including such lag effects would increase tremendously the complexity of our modelling framework and is beyond the reach of this article.

The other medium term process modelled is the relocation of a fraction of dwellers and jobs. We also stated such processes to occur over the course of 2/3 years. Here this is not a strong assumption since the dynamics of population relocation indeed occur over the course of all time steps, and it is a type of implementation already adopted in many LUTI models \cite{wegener2004land}.

Levels of stochasticity depend on the time scales involved: the smallest scales have deterministic dynamics whereas the longer exhibit randomness.

In this article, we will in the following study the coupling between the LU-EC-I sub-models and a simplified T sub-model: in an exploratory perspective we do not take into account the effect of congestion; furthermore we consider simple assumptions for the location sub-model and neglect agglomeration economies.

\subsection{Model description}

\subsubsection{Environment}

The mega-city region is modelled through a two scales spatial zoning. The model world is composed by a lattice of $K^2$ square zones, that are the basic units used to quantify land use. We assume that each zone $k$ is characterised at time $t$ by its resident workers $A_k(t)$ and number of jobs $E_k(t)$. The MCR is at a higher level decomposed into $M$ administrative areas that correspond to the city governance levels, to which are assigned $M$ abstract agents called mayors. We also introduce in the model another agent, that corresponds to a regional authority at the level of the MCR. We argue here that this is a realistic assumption in the sense that even though proper MCR scale governance bodies hardly exists beyond embryonic or informal cooperation entities, most MCR in the world are a major part of existing administrative regions, which may have strong power in terms of transport and/or urban planning \citep{xu2010governance}. For instance, Rhine-Ruhr region, in Germany, already mentioned, belong to the ``North-Rhine Westphalia'' Lander, of which it represents roughly 20\% of area, 60\% of population and 80\% of GDP.

In line with most models on residential mobility used in LUTI models, micro-economic agents will relocate in order to maximise their accessibility to amenities. Also, new transportation infrastructure decisions will be taken by governance agents based on a criteria maximising the accessibility increase in their area.

On top of this patch-level land-use and governance setup, we introduce a transportation network localised in space by its node coordinates and which links are characterised by a speed $v_G$ (ratio speed on the network versus speed on virtual local roads, not modelled here). This representation allows us to compute a shortest path distance matrix between each pair of patches $D = d_{k,k'}$.  The accessibility to jobs is then defined at patch-level as a Hansen-like accessibility function with a distance decay parameter $\lambda$ capturing typical commuting range, by:

\begin{equation}\label{eq:accessibility}
X^{(A)}_k = A_k\cdot \sum_{k'} E_{k'} \exp{\left(-\lambda \cdot d_{k,k'}\right)}
\end{equation}

In operational models, the decay parameter $\lambda$ is calibrated using commuting data with spatial interaction models \citep{fotheringham1989spatial}. In the toy-model implementation we describe below, $\lambda$ parameter value is set such that a majority of a subregion accessibility is captured within that centre.

The dynamics of the model are simulated using discrete time steps. At each time step, (i) land-use is updated (i.e. some actives and employments relocate); and (ii) the transportation network evolves. Each step is described with more details in the following.

\begin{figure}
	\includegraphics[width=\linewidth]{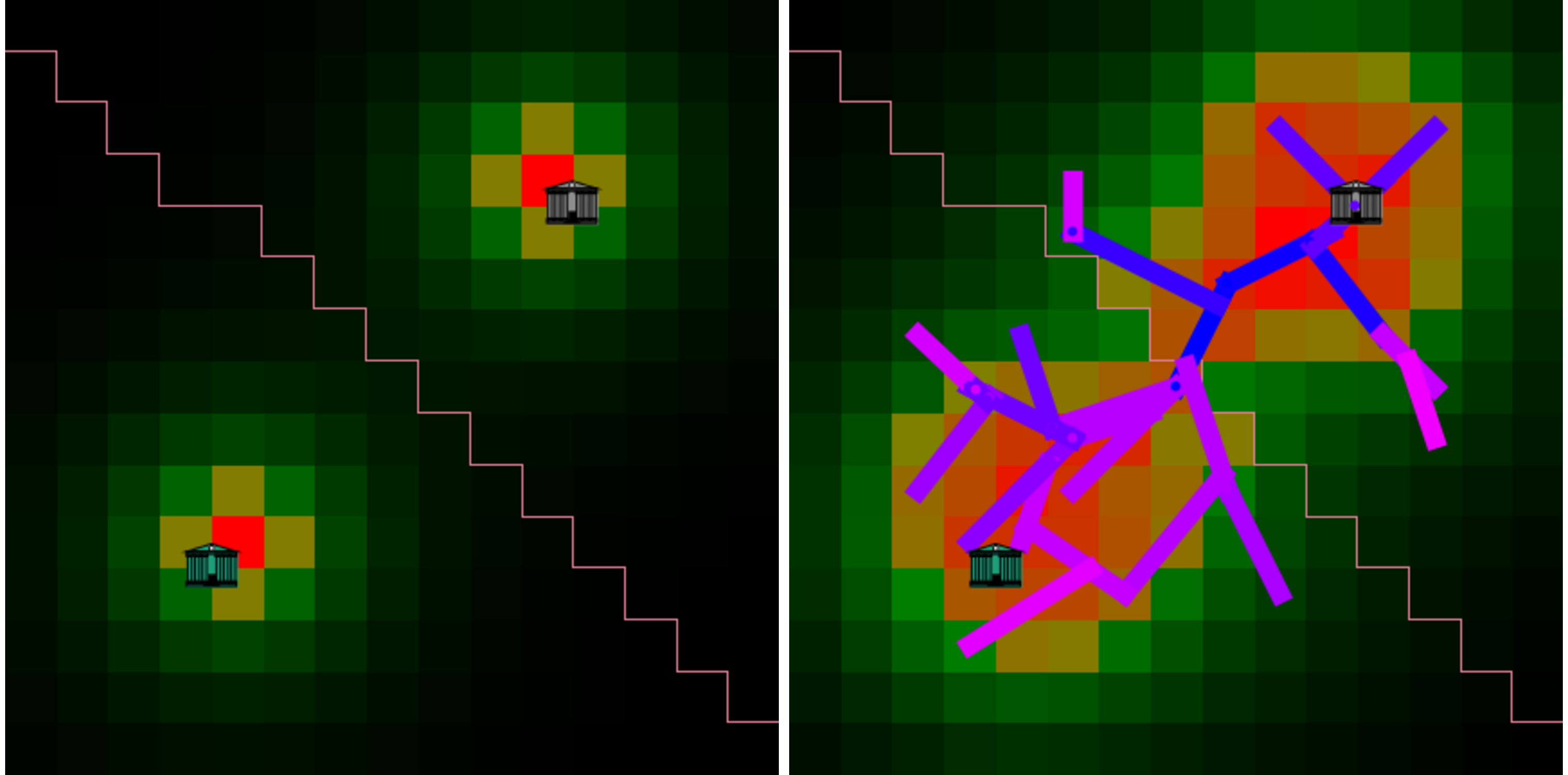}
	\caption{\textbf{Model setup with two centres.} Colour level of patch areas gives the number of actives at initial step in that simple two-centres setting, constructed with exponentially decreasing densities. Mayors are visualised at the centre of each governance zone, and the boundary between these two zones is highlighted. (\textit{Left}) Initial configuration; (\textit{Right}) Example of final configuration after several time steps.\label{fig:setup}}	
\end{figure}

The Fig.~\ref{fig:setup} illustrates a setup with two centres of equal size, and the final configuration after having evolved the land-use and the transportation network several times.

%%%%%%%%%%%%%%
%%%%%%%%%%%%%%

\subsubsection{Evolution of land-use}

For the land-use module, the evolution rules are based on the \cite{lowry1964model} model. By comparison with residents and employments relocation, the evolution of transportation infrastructure is much slower~\citep{wegener2004land}. Note that we do not consider land values, rents or transportation costs, that are the core of models in Urban Economics such as the Alonso and Fujita models for example \citep{lemoy2017exploring}.

Residents and jobs relocate given some utility functions that can depend on several characteristics of the areas. We consider simple drivers, one positively linked to a demand for a high accessibility, and one negatively linked to a demand of low density (urban sprawl). These two effects are aggregated in a simple way, by taking a Cobb-Douglas function for utilities of actives and employments, similarly to \cite{caruso2011morphological} (which is equivalent to have a linear aggregation of the logarithm of explanatory variables).

\begin{equation}\label{eq:utility}
U_k^{(A)} = {X_k^{(A)}}^{\gamma_A}\cdot {F_k^{(A)}}^{1-\gamma_A}
\end{equation}

Employments follow an analog expression with a dedicated weight parameter $\gamma_E$. The utility function is simply influenced only by accessibility and by an indicator of local urban form $F_k$ called \emph{form factor}. This expression of the utility can be interpreted as a trade-off made by agents between the maximisation of accessibility and the maximisation of a ``comfort'' function given by the form factor. Here, we assume a simple expression for it and take it for actives as $F_k^{(A)} = \frac{1}{A_k \cdot E_k}$, what means that population is repulsed by density. As we study a stylised model, we do not include more realistic expressions. The combination of the positive effect of accessibility to the negative effect of density produces a tension between contradictory objectives allowing a certain level of complexity typical of urban systems \citep{fujita1999evolution} already in the land-use sub-model alone. The form factor for employments is taken as $F_k^{(E)}=1$ for the sake of simplicity and following the rationale that jobs can aggregate far more than dwellings.

Relocations are then computed in a deterministic way, following a discrete choice model similar to \citep{ben1985discrete}, which yields the value of actives at the next step as

\begin{equation}\label{eq:discretechoicereloc}
A_i(t+1) = (1 - \alpha) \cdot A_i(t) + \alpha \cdot \left(\sum_j{A_j(t)}\right)\cdot\frac{\exp{(\beta \tilde{U}_i(A))}}{\sum_j{\exp{(\beta \tilde{U}_j(A))}}}
\end{equation}

where $\beta$ is the Discrete Choice parameter that can be interpreted as a ``level of randomness''. This parameter $\beta$ is a model parameter that can be changed in the numerical experiments. When $\beta \rightarrow 0$, all destination patches have an equal probability from any origin patch, whereas $\beta \rightarrow \infty$ yields a fully deterministic behavior towards the patch with the best utility. $\tilde{U}_i$ are the utilities normalised by the maximal utility. $\alpha$ is also a model parameter, which corresponds to the fixed fraction of actives relocating at each time step. Relocation of jobs follow again a similar expression.

\subsubsection{Network evolution: governance process}

Given a planning actor, we make a strong assumption: the transport infrastructure provision is decided so as it maximise accessibility gains for local residents (or all region for regional planning actor). Following mechanisms implemented by \cite{Xie2011} and empirical work from \cite{chen2014spatio}, which study the evolution of accessibility as a consequence of Guangzhou metro network growth, the optimization of accessibility increase seems a reasonable assumption. This is indeed a simplistic assumption since several other factors are also used to assess transport infrastructure provision (among for example time gains, security gains, reduction of greenhouse gases emissions). When accessibility gains are accounted for, they can be used to quantify employment growth and/or productivity increase through agglomeration economies, and typically represent a third of the overall investment benefit \citep{horcher2020public}. By putting emphasis on accessibility gains only, we stay in the logic of a simple ``toy'' model: this process is the most interesting in the context of our research question, since agglomeration economies operate at MCR scale \citep{boussauw2018planning}. Furthermore, accessibility is a factor for which the spatial distribution is important, and is thus central in such a spatially explicit model of land-use/network dynamics.

How is managed the power equilibrium between planning actors? Some of the models reviewed above, in particular \cite{Xie2011}, are based on game theory to model the behavior of stakeholders. This framework has already been widely applied for modelling in social and political sciences, in relation to questions dealing with cognitive interacting agents with individual interests~\citep{ordeshook1986game}. \cite{abler1977spatial} (p.~487) suggest a location decision problem for coffee farms on Kilimanjaro as a game combining a production strategy and a location strategy (fixing then the environmental conditions). \cite{batty1977game} introduces models of the planning process based on game theory. This framework has furthermore already been used in transportation investment studies, such as by~\cite{Roumboutsos2008209} which use the notion of Nash equilibrium to understand choices of public or private operators concerning the integration of their system in the broader mobility system. We will use game theory paradigms to integrate governance in a simple way in our model.

The governance part of the model has therefore the following rationale:
\begin{enumerate}
\item Two levels of governance are included, namely a central planning actor (the region, or regional government), and local planning actors.
\item The total stock of infrastructure built at one governance time step is constant, which corresponds to a steady development state, therefore before the saturation of the network.
\item Assuming a new infrastructure is to be built, the transport provision decision making process can be either from top-down decision (region) or from the bottom-up. We make the assumption that the processes behind the determination of the level of decision are exogenous and not taken into account in the model. This step is thus determined following a uniform random law. The probability of the decision to be at the regional level is therefore a model parameter that we denote $\xi$. Specifically, in the model, two new road segments of a given length are built at each time step. The length of built infrastructure segments is also a model parameter denoted $l_r$. In the case of local decision, roads are attributed successively to mayors (one road maximum per mayor) with probabilities $\xi_i$ which are proportional to the number of jobs of each subregion.
\item If the decision is taken at the local level, a virtual negotiation between planning actors occurs. Indeed, the issue of political coordination is crucial in the dynamics of city-regions \citep{scott2019city}. We assume that (i) the initiator of the new infrastructure can be any of the local planning actors, but cities with more economic activity will have more chance to build the new transportation link; (ii) negotiations for possible collaborations are only done between neighbour cities, what is related to the medium range of infrastructure segments considered; (iii) collaborations can occur between a maximum of two local planning agents. This choice is due to simplicity, as games with more players have furthermore chaotic dynamics \citep{sanders2018prevalence}. Specifically, in the model, the local planning agents, indexed by $i\in\{0,1\}$, must choose a strategy $S_i$ among: (i) not collaborating (NC), and (ii) collaborating (C). To each combination of strategies are associated utilities for each player, which correspond to optimal infrastructures, that are used to determine probabilities of each strategies as detailed below.
\end{enumerate}

\subsubsection{Evaluation of cooperation}

We detail now the way the cooperation probabilities are established. We denote $Z^{\ast}_i(S_0,S_1)$ the optimal infrastructure in terms of accessibility gain for area $i$ with $(S_0,S_1)\in \{(NC,C),(C,NC),(NC,NC)\}$ which are determined by an heuristic in each zone separately (see implementation details), and $Z^{\ast}_C$ the optimal common infrastructure computed with a two segments infrastructure on the union of both areas. It corresponds to the case where both strategies are $C$. Marginal accessibility for area $i$ and infrastructure $Z$ is defined as $\Delta X_i(Z)=X^Z_i - X_i$. We introduce construction costs, noted $I$ for a road segment, assumed spatially uniform. We furthermore introduce a cost of collaboration $J$ that corresponds to a shared cost for building a larger infrastructure.

If the outcome of strategies is $(C,C)$, $Z^{\ast}_C$ will be built. If one or both strategy is $NC$, $Z^{\ast}_i(S_0,S_1)$ are built separately, and the collaborating player loose the collaboration cost. The values of utility gains for each player and each possible decision configuration can be summarised as a payoff matrix for the game. The payoff matrix is the following, the players being written $i\in \{ 0;1\}$ (such that $1-i$ denotes the player opposed to $i$)

\medskip
\begin{center}
\begin{tabular}{ |c|c|c| } 
 \hline
 0 $|$ 1  & C & NC \\ \hline
 C & $U_i = \Delta X_i(Z^{\ast}_C) - I - \frac{J}{2}$
   & $\begin{cases}U_0 = \Delta X_0(Z^{\ast}_0)-I \\U_1 = \Delta X_1(Z^{\ast}_1)-I - \frac{J}{2}\end{cases}$ \\ \hline
 NC & $\begin{cases}U_0 =  \Delta X_0(Z^{\ast}_0)-I - \frac{J}{2}\\U_1 =\Delta X_1(Z^{\ast}_1)-I\end{cases}$
   & $U_i = \Delta X_i(Z^{\ast}_i) - I$ \\
 \hline
\end{tabular}
\end{center}
\medskip

Note that to simplify, we assumed the cost parameters $I$ and $J$ with the dimension of an accessibility. We will furthermore see that since only accessibility differentials have impact on the choices, the construction cost $I$ does not play any role. This payoff matrix is used in two games corresponding to complementary processes:
\begin{itemize}
	\item The coordination game in which players have a mixed strategy. We consider for this game the Nash equilibrium, which is a strategy point in a discrete non-collaborative game for which no player can improve his gain by changing his strategy~\citep{ordeshook1986game}. This implies a competition between players.
	\item A heuristic game according to which players take their decision following a discrete choice model. It implies only a maximisation of the utility gain and an indirect competition.
\end{itemize}

The inclusion of these two different processes is in line with a multi-modelling approach \citep{see2000hybrid}, which solves the issue of competing modelling assumptions by comparing them.

We write $p_i = \Pb{S_i = C}$ the probability of each player to collaborate. These probabilities can be computed for each game.

\paragraph{Nash equilibrium}

We solve the mixed strategy Nash Equilibrium for this coordination game in all generality. The computation is detailed in Supplementary Material. By writing $U_i(S_i,S_{1-i})$ the full payoff matrix, we have the expression of probabilities

\[
p_{1-i} = - \frac{U_i(C,NC) - U_i(NC,NC)}{\left(U_i(C,C) - U_i(NC,C)\right) - \left(U_i(C,NC) - U_i(NC,NC)\right)}
\]

What gives with the expression of utilities previously given,

\begin{equation}
p_i = \frac{J}{\Delta X_{1 - i}{Z^{\star}_{C}} - \Delta X_{1 - i}{Z^{\star}_{1 - i}}}
\end{equation}

This expression can be interpreted the following way: in this competitive game, the likelihood of a player to cooperate will decrease as the other player gain increases, and somehow counter-intuitively, will increase as collaboration cost increases.

\paragraph{Discrete choice decisions}

Using the same utility functions, a random utility model for a discrete choice also provides expressions for probabilities. We have for player $i$ the utility differential between the choice $C$ and the choice $NC$ given by

\[
U_i(C) - U_i(NC) = p_{1 - i} \left( \Delta X_{i}{Z^{\star}_{C}} - \Delta X_{i}{Z^{\star}_{i}}\right) - J
\]

Under the classical assumption of a model with a random utility where the random term follows a Gumbel law~\citep{ben1985discrete}, we have $\Pb{S_i=C} = \frac{1}{1 + \exp{[-\beta_{DC}(U_i(C) - U_i(NC))]}}$, where $\beta_{DC}$ is a discrete choice parameter. For the sake of simplicity, we fix this parameter at $\beta_{DC} = 400$ a very high value which is close to have a deterministic choice at this step.

We substitute the expression of $p_{i-1}$ in the expression of $p_i$, what leads $p_i$ to verify the following equation

\begin{equation}
p_i = \frac{1}{1 + \exp{\left(-\beta_{DC}\cdot \left(\frac{\Delta X_{i}{Z^{\star}_{C}} - \Delta X_{i}{Z^{\star}_{i}}}{1 + \exp{\left(- \beta_{DC}(p_i \cdot (\Delta X_{1 - i}(Z^{\star}_{C}) - \Delta X_{\bar{i}}(Z^{\star}_{1 - i})) - J)\right)}} - J \right)\right)}}
\end{equation}

We demonstrate (see Supplementary Material) that there always exists a solution $p_i \in [0,1]$, and we solve it numerically in the model to determine the probability to cooperate. We also exhibit in Supplementary Material the empirical behavior of $p_i$ as a function of $J$ for various values of accessibility differentials and $\beta_{DC}$. As expected, it always decreases when $J$ increases, what confirms the complementary of this game with the first Nash game. The choice $\beta_{DC} = 400$ furthermore provides a stronger non-linearity than with smaller values, what is also a complementary aspect to include.

\paragraph{Random decision}

We also consider a baseline mechanism, which does not assume negotiations, but which in the case of a local decision draws randomly a mayor, following a uniform law with probabilities proportional to the number of employments of each.

\subsection{Model implementation}

The model is implemented in \texttt{NetLogo}~\citep{wilensky1999netlogo} as this platform is particularly suited for such exploratory and interactive models. A particular care is taken for the computation of accessibility and shortest paths, as a dynamic reevaluation of network distance is necessary for each new potential infrastructure, what becomes rapidly a computational burden. We use therefore a computation of shortest paths based on dynamic programming~\citep{tretyakov2011fast}. For the determination of the optimal infrastructure, the order of magnitude of the total number of infrastructures to explore is in $O(l_r\cdot K^2)$, as $K^2$ is the number of patches and assuming that all potential infrastructures have their extremities in the centre of a patch. For each patch, we will have an infrastructure for each other patch in a radius $l_r$, what asymptotically corresponds to the perimeter of the circle $2\pi l_r$. Furthermore, as detailed in Supplementary Material, we assume a snapping heuristic to existing infrastructures to keep a consistent network. This infrastructure optimization problem considerably increases the operational computational cost, and we use an heuristic exploring a fixed number $N_I$ of randomly chosen infrastructures. In practice, we take $N_I = 50$ which gives a reasonable coverage of possible choices with a realistic computational time for each model run. Low values for $N_I$ will lead to random infrastructure choices, while high values will imply an exhaustive search, ensuring that the infrastructure segment is optimal.

The model is open-source, and its implementation is available on the open repository of the project at \url{https://github.com/JusteRaimbault/TransportationGovernance}. Simulation results used in this paper are available on the Dataverse repository at \url{https://doi.org/10.7910/DVN/V3KI2N} and our modelling framework is thus reproducible by external readers. Regarding technical requirements to run the model, no specific constraints in terms of memory or number of cores exist to simply run the model with reasonable world sizes (less than a width of 50). To run numerical experiments described below, and in particular calibration, high performance computing infrastructures are needed as a high number of model runs (often more than 10,000) are required.

All model parameters are summarised in Table~\ref{tab:lutecia:parameters}. We describe here only the parameters which have not been explicitly fixed previously, and these will be the privileged parameters on which the exploration and the application of the model will be done. We give the theoretical domains of variation and default values that were obtained with trial-and-error for eyeball-validated dynamics. For example for collaboration cost $J$, higher values than $0.005$ did not change the observed dynamics. For the accessibility range $\lambda$, a value of $0.001$ yields medium commuting ranges, with a world of size 50 and a network speed of 5. The bound $\sqrt{2}\cdot K$ for infrastructure length corresponds to the diagonal of the world.

%%%%%%%%%%%%%
\begin{table*}
\caption{\textbf{Summary of Lutecia model parameters.} We also give the corresponding processes, typical bounds of the variation range and their default values.\label{tab:lutecia:parameters}}
\centering
\begin{tabular}{|c|c|c|c|c|}
  \hline
 Sub-model & Parameter & Name & Domain & Default\\
  \hline
\multirow{5}{*}{Land-use}& $\lambda$ & Accessibility range  & $]0;1]$ & $0.001$ \\\cline{2-5}
 & $\gamma_A$ & Cobb-Douglas exponents actives & $[0;1]$ & $0.85$ \\\cline{2-5}
 & $\gamma_E$ & Cobb-Douglas exponents employments & $[0;1]$ & $0.85$ \\\cline{2-5}
 & $\beta$ & Discrete choices exponent  & $[0;+\infty]$ & $1$ \\\cline{2-5}
 & $\alpha$ & Relocation rate  & $[0;1]$ & $0.1$ \\\hline
Transport & $v_G$ & Network speed  & $[1;+\infty [$ & $5$ \\\hline
\multirow{2}{*}{Governance} & $J$ & Collaboration cost  & $[0;0.005]$ & $0.001$ \\\cline{2-5}
 & $l_r$ & Infrastructure length & $]0;\sqrt{2}\cdot K [$ & $2$ \\\hline
\end{tabular}
 \end{table*}
%%%%%%%%%%%%%

%%%%%%%%%%%%%%%%%%
\section{Model exploration and validation}
%%%%%%%%%%%%%%%%%%

In this section, we detail the validation and exploration of the theoretical model for: land-use module only (to control the sprawl speed), governance module only (to test for the influence of coordination between planning actors), and both modules activated (co-evolution), to test the condition of emergence of a coordination at MCR scale. The idea is to proceed to elementary experiments by making either land-use, or network, or both, evolve, and studying the consequences on the different aspects. 

Numerical experiments are realised using the OpenMOLE model exploration software \citep{reuillon2013openmole}, which (i) can embed any model as a black box; (ii) integrates state-of-the-art methods for the exploration, calibration, and validation of simulation models; (iii) provides a transparent access to high performance computing environments such as clusters and computation grids.

Regarding the synthetic setup for the theoretical configuration, initial distributions of actives and employments are taken around governance centres (mayors) at positions $\vec{x}_i$ using exponential kernel mixtures by

\[
A(\vec{x}) = A_{max} \cdot \exp{\left(\frac{\norm{\vec{x}-\vec{x}_i}}{r_A}\right)} ; 
E(\vec{x}) = E_{max} \cdot \exp{\left(\frac{\norm{\vec{x}-\vec{x}_i}}{r_E}\right)}
\]

This configuration with two centres of equal size is shown in left panel of Fig.~\ref{fig:setup}.

%%%%%%%%%%%%%%%%%%
\subsection{Land-use}

Land-use dynamics always converge towards an asymptotic state when the network does not evolve. We demonstrate the existence of the equilibrium in Supplementary Material. We proceed to an exploration of the behavior of the land-use model alone, i.e. making actives and employments evolve with the infrastructure being fixed. This aims at understanding the influence of land-use parameters on the produced urban form. We fix $\alpha = 1$ here to study the model in an extreme case (a real parametrisation corresponds to a smaller $\alpha = 0.1$, but we focus on the qualitative properties of the land-use model in this experiment). We sample other land-use parameters with a grid sampling. To quantify the evolution of urban form, we follow the indicators used by~\citep{le2012urban,raimbault2018calibration}, for the distribution of population and employments, in time and until the model has converged. These morphological indicators capture together features of a spatial distribution (these are the Moran index $I$, the entropy of the distribution $\mathcal{E}$, the average distance $\bar{d}$ and the rank-size slope $\gamma$).

In terms of morphological trajectories in time, we find that increasing values of $\beta$ have the tendency to make trajectories more simple. The cost of energy $\lambda$ and network shape have also a high influence on morphological trajectories when $\beta$ is low, but a negligible influence when it is high.

Studying the final configuration of urban form parameters, we observe a strong variability of forms (here in terms of dispersion) as a function of all parameters: for large $\beta$ values, complex diagrams emerge. For low $\beta$ values, we have a diagonal privileged for dispersion within concentrated configurations.

Finally, in order to understand the influence of parameters on total mobility within a complete trajectory, we have also studied the accumulated variation of actives given by $\tilde{\Delta} = \sum_t \sum_k \left|\Delta A_k (t)\right|$. We obtain that high values of $\gamma_A$, for a high $\beta$, allow to minimise the total quantity of relocations, which have a very low dependence in $\gamma_E$. It is therefore possible to optimise, even at fixed $\alpha$, the total quantity of urban sprawl.

To summarise, we obtain the following results on the behavior of the land-use sub-model.
\begin{itemize}
	\item A large diversity of morphological trajectories in time, i.e. the evolution of morphological indicators for the distribution of population and employments, is obtained by playing on parameters $\gamma_A, \gamma_E, \lambda, \beta$, and also on the structure of a static network.
	\item Similarly, these trajectories do not converge towards the same forms and we have thus a diversity of final forms obtained.
	\item We will use $\alpha$ parameter to control the speed of urban sprawl, and will typically take values around $0.1$, what corresponds to 10\% of actives relocating at each time step, i.e. on a period of the order of 5 years \citep{dieleman2000geography}.
\end{itemize}

%%%%%%%%%%%%%%%%%%
\subsection{Governance}

In order to understand the influence of governance parameters on network structure produced by the model, we proceed to a simple experiment in the case of two cities in interaction, without an initial network. Parameters for the land-use model are fixed at $\gamma_A = \gamma_E = 0.8$ (close to default values to ensure a stronger role of accessibility rather than density); $\beta = 2$ to limit dispersion; $\lambda = 0.001$ (default value); and $\alpha = 0.16$ (close to default). The length of infrastructure segments is fixed to $l_r = 2$, as small increments in the network will imply more decisions to be taken and a higher effect of governance processes. The influence of land-use evolution is not the focus of this experiment, hence their default values. We also do not focus on discrepancies between the type of game used, and only consider the discrete choice game. The reference situation is given by a fully regional decision level, corresponding to $\xi = 1$. We compare it to two situations in which the level of decision is fully local ($\xi = 0$) but for which we force the possibility of collaboration to extreme values by the intermediate of the cooperation cost, taken respectively as $J=0$ and $J=0.005$ (this parameter cannot be interpreted in terms of absolute value).

%%%%%%%%%%%%
\begin{figure}
	\includegraphics[width=\linewidth]{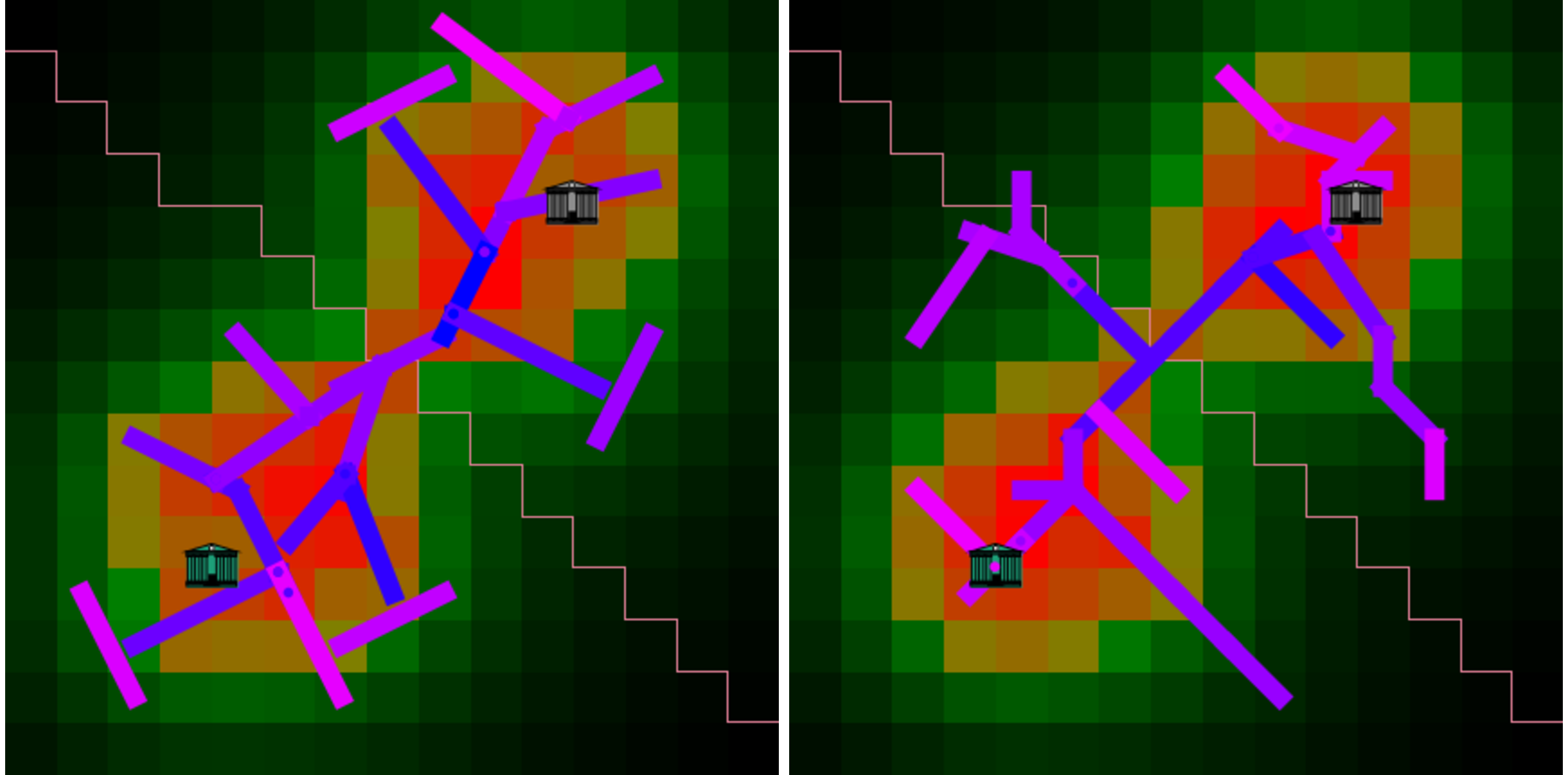}
	\caption{\textbf{Network topology obtained for different levels of governance.} The model is initialised on a symmetric synthetic configuration with two centres (Fig.~\ref{fig:setup}). Parameters for the evolution of land-use are $\gamma_A = \gamma_E = 0.8 ; \beta = 2 ; \lambda = 0.001 ; \alpha = 0.16$, and for network evolution $l_r = 2$ and a discrete choices game. The evolution is stopped at fixed stock $S = 50$ and the heuristic exploration done for $N_I = 200$. (\textit{Left}) Regional decision level with $\xi = 1$; (\textit{Right}) Local level with $\xi = 0$ and low level of collaboration, obtained with a high cost of cooperation $J=0.005$.\label{fig:lutecia:governance}}
\end{figure}
%%%%%%%%%%%%

The final configurations obtained in two settings (regional only, and local with a low level of collaboration) are shown in Fig.~\ref{fig:lutecia:governance}. Network shapes are visually different and have particular structural characteristics. In the case of the regional decision, a structuring arc links the two centres, from which extensions branch, first perpendicularly and then in parallel. The structure obtained in the case of local decisions with a low collaboration is also tree-like but has less branches, the extensions following mainly the existing branches. Also, we note that the local collaborative network, corresponding to the final configuration in Fig.~\ref{fig:setup}, seems to be less optimal in terms of covering than the two networks shown here in Fig.~\ref{fig:lutecia:governance}, but contains loops which make it more resilient. Concerning the urban structure, comparing the distance between cities in the two configurations suggests that the local level better conserves the initial structure, i.e. that regional decisions would induce more relocations. These results show, with the parameter values used for governance parameters $\xi$ and $J$, a significant impact of governance structure on MCR dynamics.

\subsection{Co-evolution}

In a last experiment, we study more directly the effect of co-evolution, in particular on land-use variables. Therefore, we consider again the previous bi-centric configuration, with a disequilibrium of population and employments between the two centres (in practice with a rate of 2), and different distances (close configuration, at a distance of $0.4 \cdot K$, and far configuration, at a distance of $K$). We fix a random local governance structure (choice of only one constructor with a probability proportional to employments) as a way not to include effects of the governance structure. We fix the land-use parameters as $\gamma_A = 0.9$ and $\gamma_E = 0.65$ (actives relocating more than employments), $\lambda = 0.005$ (relatively long range commute), $\beta = 1.8$ (fewer randomness), $\alpha = 0.1$ (standard value for relocation rate); and the network parameters as $l_r = 1, v_0 = 6$ (medium-size but fast infrastructures). We study the influence of the decision level $\xi$ on two indicators, namely: (i) the total accessibility gain between the initial and the final state, expressed as a rate $\frac{X(t_f)}{X(t_0)}$; and (ii) the evolution of relative accessibility between the two centres, given by $\frac{X_0(t_f)}{X_0(t_0)} / \frac{X_1(t_f)}{X_1(t_0)}$. The first indicators allows us to understand the global benefit, whereas the second expresses the inequality between the centres (for example, is the smallest centre drained by the main centre, or does it benefit from it). They are of particular interest regarding our general research question as they express first if the mega-city region overall benefits of the collaboration or competition between planning actors, and secondly how its polycentric structure is affected by co-evolutionary dynamics.

\begin{figure*}
	\includegraphics[width=\linewidth]{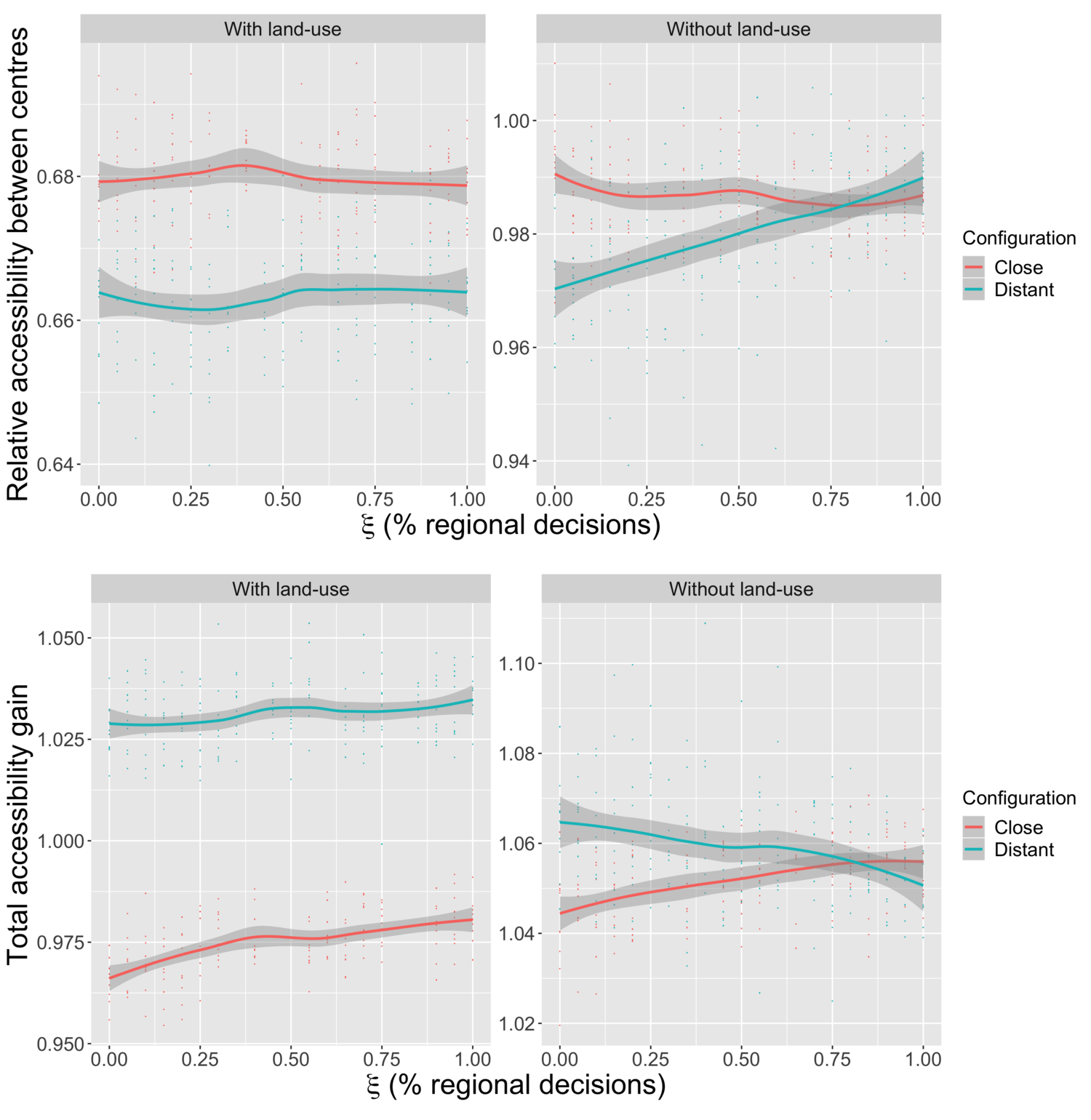}
	\caption{\textbf{Impact of co-evolution on accessibility in the Lutecia model.} We proceed to 10 repetitions with fixed parameters $\gamma_A = 0.9, \gamma_E = 0.65, \lambda = 0.005, \beta = 1.8, \alpha = 0.1, l_r = 1, v_0 = 6$, for a random local governance, and an evolution with constant stock $S=20$. We compare the evolution with network only (without land-use) and with co-evolution, for the close and distant configurations. \textit{(Top)} Evolution of the relative accessibility between centres as a function of $\xi$, with and without land-use (columns) for the two configurations (colours);  \textit{(Bottom)} Evolution of the total accessibility gain as a function of $\xi$, with and without land-use (columns) for the two configurations (colours).\label{fig:lutecia:coevol}}
\end{figure*}

Results of the experiment are given in Fig.~\ref{fig:lutecia:coevol}. Without land-use sub-module activated, the relative accessibility between centres is reduced for distant bi-centric configuration, when 100\% of decisions are at regional scale, meaning that this processes embed a potential to reduce accessibility inequalities. However, interestingly, when land-use module is activated, this effect disappears (blue curve behavior between top panels of Fig.~\ref{fig:lutecia:coevol}), illustrating once again the strong path-dependency of transport / territory co-evolution. We furthermore interpret the behavior of the accessibility gain (bottom row of Fig.~\ref{fig:lutecia:coevol}) as a direct effect of co-evolution processes: in the case of distant centres, the effect of $\xi$ on relative accessibility gains is cancelled when we add the evolution of land-use. In the case of a network evolving alone, a local decision is optimal for total accessibility, whereas in the case of a co-evolution of processes, the optimal is at a fully regional decision. We interpret this stylised fact as the existence of a need for coordination for the success of a coupled evolution of the transportation network and land-use, what can be put in correspondence with the concept of Transit Oriented Development \citep{loo2010rail}. In the case of close centres, the regional decision is always optimal, corresponding then to a more integrated metropolitan area. This is a key result regarding our research question: this last experiment suggests the existence of ``co-evolution effects'', in the emergence of a need for regional coordination in the case of a coupled evolution. There would exist a critical distance under which integrated governance at MCR level would be optimal regarding accessibility level.

%%%%%%%%%%%%
%%%%%%%%%%%%

%%%%%%%%%%%%%%%%%%
\section{Application to Pearl River Delta}
%%%%%%%%%%%%%%%%%%

In the last section of this article, we apply our modelling framework to a semi-realistic case study: the Pearl River Delta (PRD), in China. In this context, using modelling tool to assess the hypothesis of multi-level governance is relevant according to \cite{liao2017ouverture}. We call this application ``semi-realistic'' because of the stylised initial configuration of the model, with poor spatial granularity and a simplified description of major transport infrastructures only. Also, due to data available, land-use is fixed in our model, the transportation network evolving from 1990 to 2010. This indeed is an important simplification, since PRD demographics has increased over the last 20 years. However, in our modelling framework, we believe these assumption to be nevertheless acceptable, in the sense that we test for the influence of a macroscopic governance parameter only, and do not aim at reproducing the full complexity of PRD dynamics.

%%%%%%%%%%%%%%%%%%
\subsection{Description of Pearl River Delta}

The Pearl River Delta, China, is one of the most economically prosperous urban region in People's Republic of China. The mega-city region is strongly polycentric \citep{yeh2020cities} and composed by 6 major centres, including Guangzhou and Shenzhen, and is in close link with the Special Administrative Regions of Hong-Kong of Macao. Altogether, around 60 millions people live in the core of the MCR (depending on how it is delimited and how population if estimated). This region witnesses the complexity of the implications of multi-level governance on regional development: \cite{shen2002urban} provides evidences studying various factors from an empirical point of view. As recalled by~\cite{xu2005city} who study strategies of the City of Guangzhou to gain influence compared to the regional entity, the competition between cities in Pearl River Delta has always been increasing since the economic reforms (started by Deng Xiaoping in 1978). Guangzhou has thus elaborated intensive restructuring strategies with major infrastructure developments, to confirm its central place in Pearl River Delta. The development of transportation infrastructures, and in particular expressways since the 90's (expressway length raised from 15 km in 1990 to 3500 km in 2010, and 6100 km are planned for 2020), has significantly changed accessibility patterns and contributed to the emergence of the MCR~\citep{hou2011transport}. It was suggested by \cite{liao2017ouverture} that processes similar to a multi-level governance recently emerged in China, in the context of economic activities growth. We try with our model to test the relevance of this hypothesis regarding the urban structure of the MCR and the way our model better fit the existing transportation network in PRD.

%%%%%%%%%%%%%%%%%%
\subsection{Model setup and calibration procedure}

We work on a simplified raster configuration (5km cells) for population in Pearl River Delta, and on the stylised freeway network. We choose to consider only the road network since, following \cite{hou2011transport}, it has been the main driver of changes in accessibility patterns. In comparison, the railway network accelerated development is rather recent and has less contributed to accessibility changes during the timeframe we consider. Road networks are stylised from the plan given by~\cite{hou2011transport} which reproduces official documents of Guangdong province in 2010. We thus consider the freeway network in 2010 and the one planned at that time. Employment data are given for 2010 by~\cite{swerts2017database} at the level of cities. They are here uniformly distributed for each city in the simplified raster.

\begin{figure*}
    \begin{center}
	\includegraphics[width=0.9\linewidth]{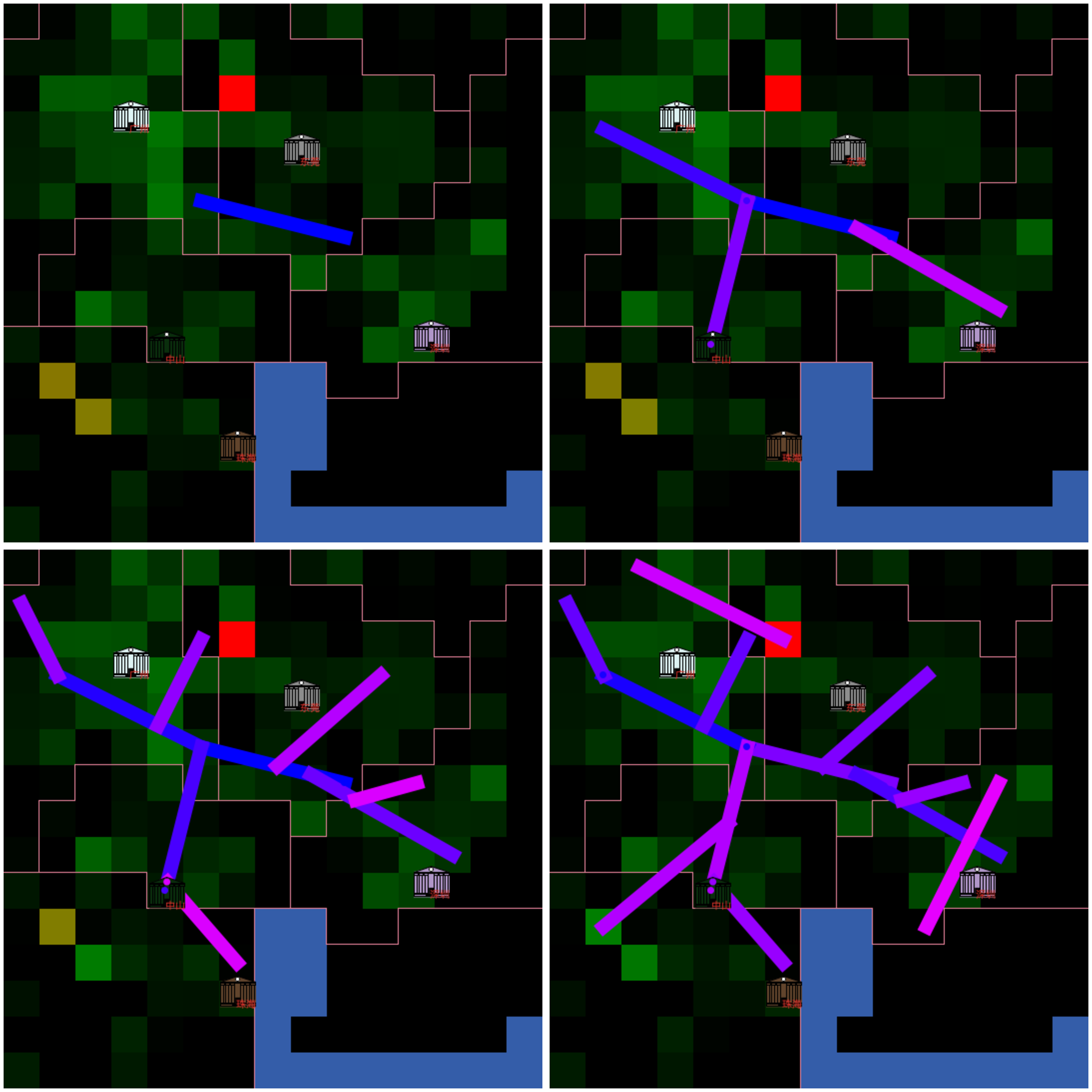}
	\end{center}
    \caption{\textbf{Example of model steps for the application to Pearl River Delta.} Land-use parameters are fixed to default values used in the application experiments, and governance parameters are $\xi = 0.9$, $J = 0.0037$ (game type is Nash). Configurations correspond, from left to right and top to bottom, to steps $t = 1, 4, 7, 10$.\label{fig:lutecia:exprd-steps}}
\end{figure*}

To illustrate network evolution in the case of the application to PRD, we show in Fig.~\ref{fig:lutecia:exprd-steps} a step-by-step application to PRD, starting with no initial network. We obtain a tree-like network which covers the region in the same way compared to the existing network but which does not includes loops. We give in Table~\ref{tab:lutecia:example} values of accessibility and travel times. With the construction of the network, travel times are reduced by one third while accessibility increases significantly given that land-use does not evolves.

%%%%%%%%%%%%%%%
\begin{table*}[h!]
\caption{\textbf{Value of indicators for the application example shown in Fig~\ref{fig:lutecia:exprd-steps}.}\label{tab:lutecia:example}}
\centering
\begin{tabular}{|c|c|c|}
  \hline
 Time step & Accessibility & Average travel time \\
  \hline
  0 & 0.856 & 31.66 \\
  3 & 0.884 & 25.01 \\
  6 & 0.898 & 21.78 \\
  9 & 0.908  & 19.54 \\\hline
\end{tabular}
\end{table*}
%%%%%%%%%%%%%%%

%%%%%%%%%%%%%%%
%%%%%%%%%%%%%%%

\begin{figure*}
	\includegraphics[width=\linewidth]{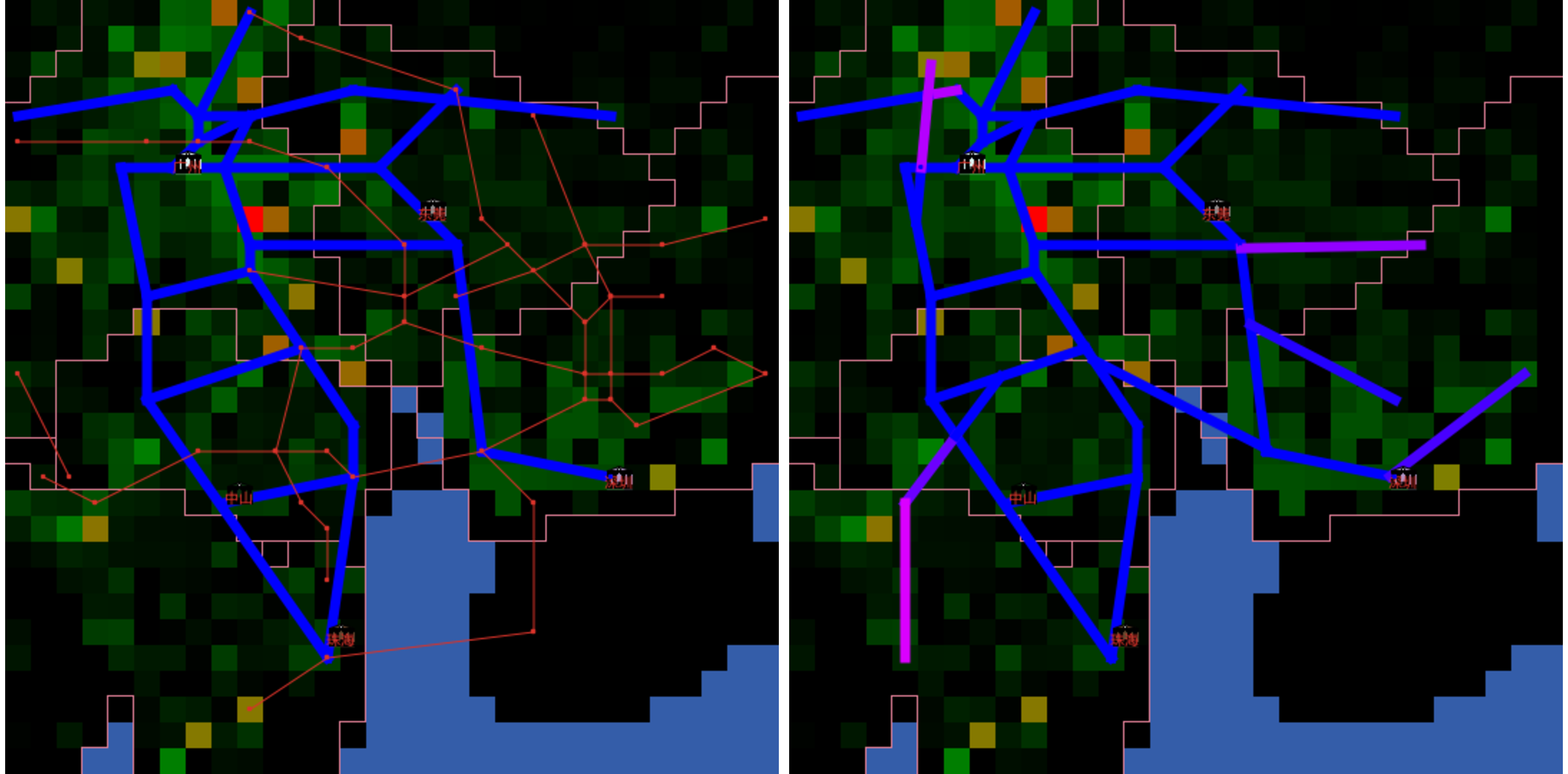}
	\caption{\textbf{Setup on real data used for model application.} (\textit{Left}) Initial networks, in blue the initial network corresponding to the network in 2010, in red and tight the planned network. (\textit{Right}) Result obtained with $\alpha = 0$ at $t_f = 7$ after a setup with the existing network; the Shenzhen-Zhongshan bridge and East developments of the network are realised in a way similar to the planned network.\label{fig:app:lutecia:realsetup}}
\end{figure*}

We then show in Fig.~\ref{fig:app:lutecia:realsetup} the population distribution and networks on which experiments on real data for the network are done in the following. The loops are realised as an extension of the existing network, and would be realised here on longer time scales in comparison to the previous example.

%%%%%%%%%%%%%
%%%%%%%%%%%%%

We now turn to the application of the model with an existing initial network. To apply such a complex model to a semi-realistic situation, one must be particularly careful. It is important to choose the adequate processes to implement and to choose the level of granularity to reproduce. We propose thus to ``calibrate'' on the shape of a given infrastructure, in the sense of determining parameter configurations for which in probability the successive built pieces of infrastructure are the closest to pieces of the target infrastructure. To calibrate on the network produced by the simulation, it must be compared to a reference network. This is however a difficult problem, as different proximity measures between networks with different significations can be used. Geometrical measures focus on the spatial proximity of networks. For a network $(E,V)=((e_j),(v_i))$, a node-based distance is given by $\sum_{i \neq i'} d^2 \left(v_i,v_{i'}\right)$. A more accurate measure which is not biased by intermediate nodes is given by the accumulated area between each pair of edges $\sum_{j \neq j'} A \left(e_j,e_{j'}\right)$ (not a distance in the proper sense) where $A(e,e')$ is the area of the closed polygon formed by joining link extremities. We consider the latest for the calibration.

%%%%%%%%%%%%%%%%%%
\subsection{Results}

We make governance parameters vary, including the type of game, with a fixed $l_r = 2$ and $\alpha = 1$ to simulate the rather long time scales needed to obtain the whole networks. We run the model on a Latin Hypercube Sampling of 4000 points in this parameter space, with 10 replications of the model for each point. Such a sampling allows to efficiently fill a high-dimensional parameter space with a fixed computational budget. The two experiments we performed correspond to different target configurations: (i) no initial network and the 2010 network as a target, in the spirit of extrapolating the most probable governance configuration which led to the current configuration; (ii) initial network as the 2010 network, and planned network as target.

\begin{figure*}
	\includegraphics[width=\linewidth]{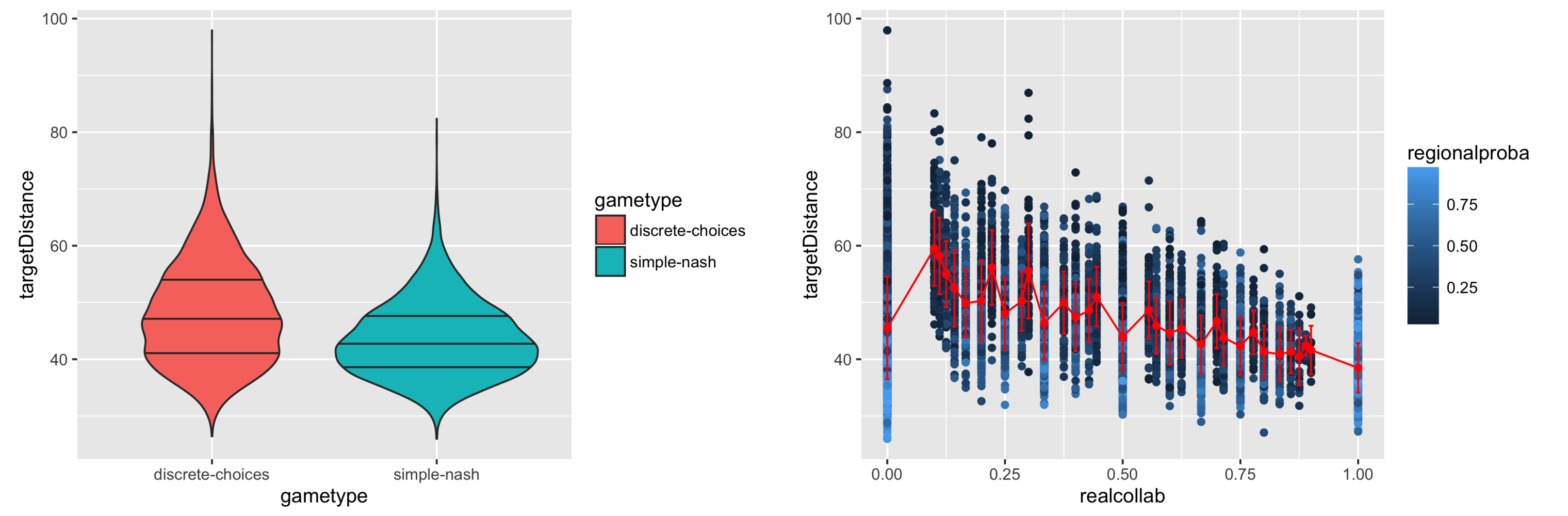}
    \caption{\textbf{Model calibration with fixed land use.} We take $\alpha = 0$ to make only network evolve, and sample the governance parameters space. (\textit{Left}) Statistical distribution of distance as a function of the type of game, in the case of the existing network as calibration target; (\textit{Right}) Distance $d_A$ to the target network, as a function of the observed collaboration probability (\texttt{realcollab}); the red curve gives the averages with standard errors.  \label{fig:lutecia:calib}}
\end{figure*}

We obtain qualitatively similar results for the two experiments, suggesting that there was no transition in the type of governance between the past network and the future network. Results are illustrated for the first one in Fig.~\ref{fig:lutecia:calib}. We obtain, by studying the graph of $d_A$ as a function of $\xi$, that the regional level is the most realistic to reproduce network shape. However, discrete choices and Nash games have a different behavior, and the Nash game is the closest to reality when $\xi$ decreases. This means that the relations between local planning actors would potentially correspond more to coordination than purely competition. When we study the variation of distance as a function of the observed collaboration level (right plot, Fig.~\ref{fig:lutecia:calib}), we obtain an interesting inverted U-shape, i.e. that the most likely configurations are the ones where there is only collaboration, or the ones where there is no collaboration at all, but no intermediate situations. Finally, the comparison of statistical distributions of distances between target configurations and the types of games shows that the difference between the games is significant only for the real network but not for the planned network (what remains a conclusion difficult to interpret).

%%%%%%%%%%%%%%%
%%%%%%%%%%%%%%%

We thus summarise from this experiment the following conclusions:

\begin{itemize}
	\item Coordination between planning actors tend to produce network configurations closer to the observations than a selfish behavior in the case of local decisions, since the Nash game give better performances than the discrete choices for low values of $\xi$ (left, Fig.~\ref{fig:lutecia:calib}).
	\item Collaboration compromises correspond to less probable networks than situations with full collaboration or with no collaboration. This could be due to the fact that the real network is rather optimal, and as suggested by the stylised exploration of the governance sub-model before, intermediate cooperation levels seem to give the less optimal networks.
\end{itemize}

These conclusions can be put into perspective with the increased competition between territorial stakeholders within the Delta revealed by~\cite{xu2005city}. Thus, this application of the model allows us to indirectly infer governance processes, by extracting through model calibration the parameter values producing network shapes that are the closest to real networks and interpreting relatively the corresponding values.

%%%%%%%%%%%%%%%
\section{Discussion}
%%%%%%%%%%%%%%%

\subsection{About the conditions for emergence of integrated governance at MCR scale}

We have suggested a stylised model aiming at a multi-scale integration of co-evolution processes, at MCR scale. The Lutecia model is theoretically validated and empirically partially calibrated for the case of Pearl River Delta (China). It allows us to test for the impacts of changes in governance structure on the spatial organisation emerging from transport/land-use interactions. This goes along with \cite{janssen2011rethinking} plea for a reconfiguration of governance structures within mega-city regions. In a few cases polycentric metropolitan area were able to implement innovative governance systems (Stuttgart, Germany or Montreal, Canada, among famous examples). However such strong cooperation systems could not be observed in polycentric MCR such as Randstad, Holland \citep{cowell2010polycentric} or Rhine-Ruhr, Germany \citep{lenechet2017peupler}. As \cite{neuman2009futures} point out, the sustainability of these MCR will be closely linked to their ability to learn new governance schemes, in the sense of an increased adaptability and flexibility of governance processes. For example, \cite{innes2010strategies} use two case studies on water planning in California to show that self-organization and collaborative dialog between local planning actors is a viable option for governance in MCR.

MCR and their transportation network structure are of interest for the policy makers for the several reasons, including (i) the world demographic growth creating need for 1.7 billion more urban at horizon 2040 in comparison to 2015~\citep{desa2014world}, especially in medium sized cities in interaction with national and international flows; (ii) transport infrastructure provision at MCR scale might increase exchanges between cities within MCR and produces macroscopic effects such as fostering the economic productivity of the region \citep{glaeser2001cities}; (iii) recent trends in mobility in term of distance and diversity of individual routines creates increasing gaps between the actions on sustainable mobility at metropolitan level and the reality of daily and non-daily flows \citep{holz2014travel,vincent2016determinants,conti2018modal}. Planning actions at this inter-urban scale are therefore crucial to foster an increase in sustainability of the mobility system.

What are the spatial factors explaining the emergence of strong regional cooperation in polycentric MCR? One relevant application of this modelling framework would be the study the assets and drawbacks of the emergence of integrated governance at MCR scale with respect to spatial and social inequalities of accessibility, for instance. An evidence-based approach to this question often rely on sparse and contextualised knowledge of recent experiments in real situations, while we provide here a simplified but integrated modelling framework to assess the possible interest for urban planning agents to collaborate at MCR scale. By focusing on the possible consequences of the level of decision for infrastructure provision, we try to help disentangle this aspect of urban complexity.

\subsection{About modelling co-evolution for multi-scalar geographical objects}

We believe our contribution to be significant on the following points: (i) to the best of our knowledge, we introduce the first model of co-evolution of land-use and transportation network with endogenous multi-level governance processes specifically tailored to understand transportation provision governance within MCR and (ii) this model is applied with a realistic geographical setting to study network growth in Pearl River Delta, China.

Our model can also be seen as a proof-of-concept that integrated transport provision / LUTI models are possible, and potentially useful to tackle the research objectives formulated by \cite{timmermans2003saga}. Thus, our contribution to the land-use-transport interaction literature is not ``from the inside'' (for example by refining accessibility measures or the description of activities in a LUTI model), but rather ``from the outside'' by embedding the LUTI framework into a broader theoretical context of co-evolution \cite{raimbault:tel-01857741} has explored simple co-evolution models for territories and networks and confirmed the potentiality of this viewpoint to understand the intricate relations between these territorial components. Let us point an original feature of this modelling framework, which considers MCR scale beyond traditional metropolitan scale which is adopted by the vast majority of LUTI models, largely ignoring the existence of border effects \citep{thomas2018city}.

Note that the approach we adopted, at this theoretical level, could be applied indifferently to public transport provision or motorway provision. Indeed, this would imply adjusting the order of magnitude of construction costs and collaboration costs, which are at this stage unique for the whole simulation, meaning only one transportation mode can be studied at a time. This would be one of the many refinements to achieve before being able to use such modelling framework in an operational context, aside more generally to include features of most LUTI models: detailed geography, role of land developers in the fabric of the city, detailed activity programs of residents, for instance. Our model being a ``toy'' model, it has not been tailored to include all these realistic aspects of the dynamics of a city.

Indeed, the way power relations are addressed in the model are simple, with a macroscopic parameter and coordination arising from theoretical game theory equations. In empirical world indeed, opportunities and individual trajectories of key stakeholders play an important role to such decisions aside the institutional willingness to cooperate \citep{ollivier2013instauration}: we believe this simplification is acceptable since our model represent long term dynamics, those perturbations being smoothed over time.

Despite the numerous simplifications operated in the model, we also suggest that it can be used to test various hypotheses about how territorial stakeholder interact. In the empirical field in particular we believe the growing number of research focusing on the cooperation between territorial stakeholders will help to implement more realistic behaviour of collective agents in such kind of models. Furthermore, our contribution opens new perspectives towards systematic testing of governance structures: for instance it can be used to assess the spatial organization at MCR level that are the most likely to be associated with the emergence of an integrated governance, following work by \cite{nechet2019modelling}, and in line with the work of \cite{nowak1994spatial} on stable cooperation configurations in game theory.

A fundamental limitation to such developments however remains the sparsity of spatial and dynamical databases covering both land-use and the evolution of transportation networks. Historical land-use data are already widely used to calibrate LUTI and land-use change models, but they generally cover shorter time spans than required to grasp the evolution of infrastructure. The construction of open and harmonised databases for the dynamics of transportation networks on long time scales is still an ongoing research issue.

Finally, the approach we adopted, using intensive computation and model simulation, also shed a light on more general issues of model validation for interaction models between transport and territories. Crucial research directions in this field indeed include (i) a more systematic benchmark and validation of existing models, in the spirit of \cite{bonnel2014survey}; (ii) the development of new methods to evaluate and calibrate interaction models, since existing examples such as methods developed by \cite{capelle2017development} rely on model-specific techniques. On the contrary, methods included in the OpenMOLE model exploration software \citep{reuillon2013openmole} are generic and provide a new type of knowledge on complex simulation models.

%%%%%%%%%%%%%%%%%%%%
%% Biblio
%%%%%%%%%%%%%%%%%%%%

%\bibliographystyle{apalike}
%\bibliography{biblio}

\begin{thebibliography}{}

\bibitem[Abler et~al., 1977]{abler1977spatial}
Abler, R., Adams, J.~S., and Gould, P. (1977).
\newblock {\em Spatial organization}.
\newblock Prentice-Hall.

\bibitem[Acheampong and Silva, 2015]{acheampong2015land}
Acheampong, R.~A. and Silva, E.~A. (2015).
\newblock Land use--transport interaction modeling: A review of the literature
  and future research directions.
\newblock {\em Journal of Transport and Land use}, 8(3):11--38.

\bibitem[Adelt et~al., 2018]{adelt2018simulation}
Adelt, F., Weyer, J., Hoffmann, S., and Ihrig, A. (2018).
\newblock Simulation of the governance of complex systems (simco): basic
  concepts and experiments on urban transportation.
\newblock {\em Journal of Artificial Societies and Social Simulation}, 21(2).

\bibitem[Batty, 1977]{batty1977game}
Batty, S.~E. (1977).
\newblock Game-theoretic approaches to urban planning and design.
\newblock {\em Environment and Planning B: Planning and Design}, 4(2):211--239.

\bibitem[Ben-Akiva and Lerman, 1985]{ben1985discrete}
Ben-Akiva, M.~E. and Lerman, S.~R. (1985).
\newblock {\em Discrete choice analysis: theory and application to travel
  demand}, volume~9.
\newblock MIT press.

\bibitem[Bonnafous, 1996]{bonnafous1996systeme}
Bonnafous, A. (1996).
\newblock Le syst{\`e}me des transports urbains.
\newblock {\em Economie et statistique}, 294(1):99--108.

\bibitem[Bonnel et~al., 2014]{bonnel2014survey}
Bonnel, P., Coulombel, N., Prados, E., Sturm, P., Arnaud, E., Boittin, C.,
  Bouzouina, L., Delgado, J.~C., Capelle, T., Delons, J., et~al. (2014).
\newblock A survey on the calibration and validation of integrated land use and
  transportation models.
\newblock In {\em Symposium" Towards integrated modelling of urban systems"}.

\bibitem[Boussauw et~al., 2018]{boussauw2018planning}
Boussauw, K., Van~Meeteren, M., Sansen, J., Meijers, E., Storme, T., Louw, E.,
  Derudder, B., and Witlox, F. (2018).
\newblock Planning for agglomeration economies in a polycentric region:
  Envisioning an efficient metropolitan core area in flanders.
\newblock {\em European Journal of Spatial Development}.

\bibitem[Bretagnolle, 2003]{bretagnolle2003vitesse}
Bretagnolle, A. (2003).
\newblock Vitesse et processus de s{\'e}lection hi{\'e}rarchique dans le
  syst{\`e}me des villes fran{\c{c}}aises.

\bibitem[Bretagnolle, 2009]{bretagnolle:tel-00459720}
Bretagnolle, A. (2009).
\newblock {\em {Villes et r{\'e}seaux de transport : des interactions dans la
  longue dur{\'e}e, France, Europe, {\'E}tats-Unis}}.
\newblock Hdr, Universit{\'e} Panth{\'e}on-Sorbonne - Paris I.

\bibitem[Capelle, 2017]{capelle2017development}
Capelle, T. (2017).
\newblock {\em Development of optimisation methods for land-use and
  transportation models}.
\newblock PhD thesis, INRIA.

\bibitem[Caruso et~al., 2011]{caruso2011morphological}
Caruso, G., Vuidel, G., Cavailh{\`e}s, J., Frankhauser, P., Peeters, D., and
  Thomas, I. (2011).
\newblock Morphological similarities between dbm and a microeconomic model of
  sprawl.
\newblock {\em Journal of geographical systems}, 13(1):31--48.

\bibitem[Cascetta et~al., 2011]{cascetta2011analysis}
Cascetta, E., Papola, A., Pagliara, F., and Marzano, V. (2011).
\newblock Analysis of mobility impacts of the high speed rome--naples rail link
  using withinday dynamic mode service choice models.
\newblock {\em Journal of Transport Geography}, 19(4):635--643.

\bibitem[Chen et~al., 2014]{chen2014spatio}
Chen, S., Claramunt, C., and Ray, C. (2014).
\newblock A spatio-temporal modelling approach for the study of the
  connectivity and accessibility of the guangzhou metropolitan network.
\newblock {\em Journal of Transport Geography}, 36:12--23.

\bibitem[Conti, 2018]{conti2018modal}
Conti, B. (2018).
\newblock Modal shift and interurban mobility: Environmentally positive,
  socially regressive.
\newblock {\em Journal of Transport Geography}, 69:234--241.

\bibitem[Coppola et~al., 2013]{coppola2013luti}
Coppola, P., Ibeas, {\'A}., dell’Olio, L., and Cordera, R. (2013).
\newblock Luti model for the metropolitan area of santander.
\newblock {\em Journal of Urban Planning and Development}, 139(3):153--165.

\bibitem[Courtat et~al., 2011]{courtat2011mathematics}
Courtat, T., Gloaguen, C., and Douady, S. (2011).
\newblock Mathematics and morphogenesis of cities: A geometrical approach.
\newblock {\em Physical Review E}, 83(3):036106.

\bibitem[Cowell, 2010]{cowell2010polycentric}
Cowell, M. (2010).
\newblock Polycentric regions: comparing complementarity and institutional
  governance in the san francisco bay area, the randstad and emilia-romagna.
\newblock {\em Urban Studies}, 47(5):945--965.

\bibitem[Desa et~al., 2014]{desa2014world}
Desa, U. et~al. (2014).
\newblock World urbanization prospects, the 2011 revision.
\newblock {\em Population Division, Department of Economic and Social Affairs,
  United Nations Secretariat}.

\bibitem[Dieleman et~al., 2000]{dieleman2000geography}
Dieleman, F.~M., Clark, W.~A., and Deurloo, M.~C. (2000).
\newblock The geography of residential turnover in twenty-seven large us
  metropolitan housing markets, 1985-95.
\newblock {\em Urban studies}, 37(2):223--245.

\bibitem[Evers and de~Vries, 2013]{evers2013explaining}
Evers, D. and de~Vries, J. (2013).
\newblock Explaining governance in five mega-city regions: rethinking the role
  of hierarchy and government.
\newblock {\em European planning studies}, 21(4):536--555.

\bibitem[Farrington, 2007]{farrington2007new}
Farrington, J.~H. (2007).
\newblock The new narrative of accessibility: its potential contribution to
  discourses in (transport) geography.
\newblock {\em Journal of Transport Geography}, 15(5):319--330.

\bibitem[Fotheringham and O'Kelly, 1989]{fotheringham1989spatial}
Fotheringham, A.~S. and O'Kelly, M.~E. (1989).
\newblock {\em Spatial interaction models: formulations and applications},
  volume~1.
\newblock Kluwer Academic Publishers Dordrecht.

\bibitem[Fujita et~al., 1999]{fujita1999evolution}
Fujita, M., Krugman, P., and Mori, T. (1999).
\newblock On the evolution of hierarchical urban systems.
\newblock {\em European Economic Review}, 43(2):209--251.

\bibitem[Glaeser and Mare, 2001]{glaeser2001cities}
Glaeser, E.~L. and Mare, D.~C. (2001).
\newblock Cities and skills.
\newblock {\em Journal of labor economics}, 19(2):316--342.

\bibitem[Gottmann, 1964]{gottmann1964megalopolis}
Gottmann, J. (1964).
\newblock {\em Megalopolis: the urbanized northeastern seaboard of the United
  States}.
\newblock MIT Press Cambridge, MA.

\bibitem[Hall and Pain, 2006]{hall2006polycentric}
Hall, P.~G. and Pain, K. (2006).
\newblock {\em The polycentric metropolis: learning from mega-city regions in
  Europe}.
\newblock Routledge.

\bibitem[Heeg et~al., 2003]{heeg2003metropolitan}
Heeg, S., Klagge, B., and Ossenbru{\"u}gge, J. (2003).
\newblock Metropolitan cooperation in europe: Theoretical issues and
  perspectives for urban networking 1.
\newblock {\em European Planning Studies}, 11(2):139--153.

\bibitem[Holz-Rau et~al., 2014]{holz2014travel}
Holz-Rau, C., Scheiner, J., and Sicks, K. (2014).
\newblock Travel distances in daily travel and long-distance travel: what role
  is played by urban form?
\newblock {\em Environment and Planning A}, 46(2):488--507.

\bibitem[H{\"o}rcher et~al., 2020]{horcher2020public}
H{\"o}rcher, D., De~Borger, B., Seifu, W., and Graham, D.~J. (2020).
\newblock Public transport provision under agglomeration economies.
\newblock {\em Regional Science and Urban Economics}, 81:103503.

\bibitem[Hou and Li, 2011]{hou2011transport}
Hou, Q. and Li, S.-M. (2011).
\newblock Transport infrastructure development and changing spatial
  accessibility in the greater pearl river delta, china, 1990--2020.
\newblock {\em Journal of Transport Geography}, 19(6):1350--1360.

\bibitem[Innes et~al., 2010]{innes2010strategies}
Innes, J.~E., Booher, D.~E., and Di~Vittorio, S. (2010).
\newblock Strategies for megaregion governance: Collaborative dialogue,
  networks, and self-organization.
\newblock {\em Journal of the American Planning Association}, 77(1):55--67.

\bibitem[Jacobs-Crisioni and Koopmans, 2016]{jacobs2016transport}
Jacobs-Crisioni, C. and Koopmans, C.~C. (2016).
\newblock Transport link scanner: simulating geographic transport network
  expansion through individual investments.
\newblock {\em Journal of Geographical Systems}, 18(3):265--301.

\bibitem[Janssen-Jansen and Hutton, 2011]{janssen2011rethinking}
Janssen-Jansen, L.~B. and Hutton, T.~A. (2011).
\newblock Rethinking the metropolis: Reconfiguring the governance structures of
  the twenty-first-century city-region.
\newblock {\em International Planning Studies}, 16(3):201--215.

\bibitem[Le~N{\'e}chet, 2010]{le2010approche}
Le~N{\'e}chet, F. (2010).
\newblock {\em Approche multiscalaire des liens entre mobilit{\'e} quotidienne,
  morphologie et soutenabilit{\'e} des m{\'e}tropoles europ{\'e}ennes: cas de
  Paris et de la r{\'e}gion Rhin-Ruhr}.
\newblock PhD thesis, Universit{\'e} Paris-Est.

\bibitem[Le~N{\'e}chet, 2011]{lenechet:halshs-00674059}
Le~N{\'e}chet, F. (2011).
\newblock {Urban dynamics modelling with endogeneous transport infrastructures,
  in a polycentric region}.
\newblock In {\em {17th European Colloquium on Quantitative and Theoretical
  Geography}}, Ath{\`e}nes, Greece.

\bibitem[Le~N{\'e}chet, 2012]{le2012urban}
Le~N{\'e}chet, F. (2012).
\newblock Urban spatial structure, daily mobility and energy consumption: a
  study of 34 european cities.
\newblock {\em Cybergeo: European Journal of Geography}.

\bibitem[Le~N{\'e}chet, 2017]{lenechet2017peupler}
Le~N{\'e}chet, F. (2017).
\newblock De l'{\'e}talement urbain aux r{\'e}gions m{\'e}tropolitaines
  polycentriques : formes de fonctionnement et formes de gouvernance.
\newblock In {\em {Peupler la terre - De la pr{\'e}histoire {\`a} l'{\`e}re des
  m{\'e}tropoles}}. Presses Universitaires Francois Rabelais.

\bibitem[Le~N{\'e}chet, 2019]{nechet2019modelling}
Le~N{\'e}chet, F. (2019).
\newblock Modelling transport provision in a polycentric mega city region.
\newblock {\em arXiv preprint arXiv:1909.02396}.

\bibitem[Lemoy et~al., 2017]{lemoy2017exploring}
Lemoy, R., Raux, C., and Jensen, P. (2017).
\newblock Exploring the polycentric city with multi-worker households: an
  agent-based microeconomic model.
\newblock {\em Computers, Environment and Urban Systems}, 62:64--73.

\bibitem[Leurent and Boujnah, 2014]{leurent2014user}
Leurent, F. and Boujnah, H. (2014).
\newblock A user equilibrium, traffic assignment model of network route and
  parking lot choice, with search circuits and cruising flows.
\newblock {\em Transportation Research Part C: Emerging Technologies},
  47:28--46.

\bibitem[Levinson, 2011]{levinson2011coevolution}
Levinson, D. (2011).
\newblock The coevolution of transport and land use: An introduction to the
  special issue and an outline of a research agenda.
\newblock {\em Journal of Transport and Land Use}, 4(2):1--3.

\bibitem[Li et~al., 2016]{li2016integrated}
Li, T., Wu, J., Sun, H., and Gao, Z. (2016).
\newblock Integrated co-evolution model of land use and traffic network design.
\newblock {\em Networks and Spatial Economics}, 16(2):579--603.

\bibitem[Liao and Gaudin, 2017]{liao2017ouverture}
Liao, L. and Gaudin, J.~P. (2017).
\newblock L’ouverture au march{\'e} en chine (ann{\'e}es 1980-2000) et le
  d{\'e}veloppement {\'e}conomique local: une forme de gouvernance
  multi-niveaux?
\newblock {\em Cybergeo: European Journal of Geography}.

\bibitem[Loo et~al., 2010]{loo2010rail}
Loo, B.~P., Chen, C., and Chan, E.~T. (2010).
\newblock Rail-based transit-oriented development: lessons from new york city
  and hong kong.
\newblock {\em Landscape and Urban Planning}, 97(3):202--212.

\bibitem[Lopes et~al., 2019]{lopes2019luti}
Lopes, A.~S., Loureiro, C. F.~G., and Van~Wee, B. (2019).
\newblock Luti operational models review based on the proposition of an a
  priori aluti conceptual model.
\newblock {\em Transport reviews}, 39(2):204--225.

\bibitem[Lowry, 1964]{lowry1964model}
Lowry, I.~S. (1964).
\newblock {\em A model of metropolis}.
\newblock Rand Corporation Santa Monica, CA.

\bibitem[Matthiessen, 2005]{matthiessen2005oresund}
Matthiessen, C.~W. (2005).
\newblock The {\"o}resund area: Pre-and post-bridge cross-border functional
  integration: the bi-national regional question.
\newblock {\em GeoJournal}, 61(1):31--39.

\bibitem[Mimeur and Th{\'e}venin, 2020]{mimeur2020analyse}
Mimeur, C. and Th{\'e}venin, T. (2020).
\newblock Analyse diachronique de la croissance du r{\'e}seau ferroviaire
  fran{\c{c}}ais entre 1860 et 1930: entre expansion connexionniste et
  s{\'e}lection hi{\'e}rarchique?
\newblock {\em Flux}, (4):69--87.

\bibitem[Murphy, 2015]{murphy2015human}
Murphy, J.~T. (2015).
\newblock Human geography and socio-technical transition studies: Promising
  intersections.
\newblock {\em Environmental innovation and societal transitions}, 17:73--91.

\bibitem[Neuman and Hull, 2009]{neuman2009futures}
Neuman, M. and Hull, A. (2009).
\newblock The futures of the city region.
\newblock {\em Regional Studies}, 43(6):777--787.

\bibitem[Nowak et~al., 1994]{nowak1994spatial}
Nowak, M.~A., Bonhoeffer, S., and May, R.~M. (1994).
\newblock Spatial games and the maintenance of cooperation.
\newblock {\em Proceedings of the National Academy of Sciences},
  91(11):4877--4881.

\bibitem[Olesen and Metzger, 2017]{olesen2017region}
Olesen, K. and Metzger, J. (2017).
\newblock The region is dead, long live the region: the {\o}resund region 15
  years after the bridge.
\newblock {\em Situated practices of strategic planning: An international
  perspective}, pages 67--83.

\bibitem[Ollivier-Trigalo, 2013]{ollivier2013instauration}
Ollivier-Trigalo, M. (2013).
\newblock L’instauration d’une {\'e}cotaxe sur les poids lourds en france:
  endurance technico-{\'e}conomique et impulsions politiques.
\newblock {\em D{\'e}veloppement durable et territoires. {\'E}conomie,
  g{\'e}ographie, politique, droit, sociologie}, 4(3).

\bibitem[Ordeshook, 1986]{ordeshook1986game}
Ordeshook, P.~C. (1986).
\newblock {\em Game theory and political theory: An introduction}.
\newblock Cambridge University Press.

\bibitem[Pagliara et~al., 2012]{pagliara2012megacities}
Pagliara, F., de~Abreu~e Silva, J., Sussman, J.~M., and Stein, N. (2012).
\newblock Megacities and high speed rail systems: which comes first?

\bibitem[Pemberton, 2000]{pemberton2000institutional}
Pemberton, S. (2000).
\newblock Institutional governance, scale and transport policy--lessons from
  tyne and wear.
\newblock {\em Journal of Transport Geography}, 8(4):295--308.

\bibitem[Raimbault, 2017]{raimbault2017models}
Raimbault, J. (2017).
\newblock Models coupling urban growth and transportation network growth: An
  algorithmic systematic review approach.
\newblock {\em Plurimondi}, (17).

\bibitem[Raimbault, 2018a]{raimbault2018calibration}
Raimbault, J. (2018a).
\newblock Calibration of a density-based model of urban morphogenesis.
\newblock {\em PloS one}, 13(9):e0203516.

\bibitem[Raimbault, 2018b]{raimbault:tel-01857741}
Raimbault, J. (2018b).
\newblock {\em {Characterizing and modeling the co-evolution of transportation
  networks and territories}}.
\newblock Theses, {Universit{\'e} Paris 7 Denis Diderot}.

\bibitem[Reuillon et~al., 2013]{reuillon2013openmole}
Reuillon, R., Leclaire, M., and Rey-Coyrehourcq, S. (2013).
\newblock Openmole, a workflow engine specifically tailored for the distributed
  exploration of simulation models.
\newblock {\em Future Generation Computer Systems}, 29(8):1981--1990.

\bibitem[Ribeill, 1985]{ribeill1985aspects}
Ribeill, G. (1985).
\newblock Aspects du développement du réseau ferré français sur la longue durée. L'approche historique.
\newblock {\em FLUX Cahiers scientifiques internationaux Réseaux et Territoires}, 1(1):10--25.

\bibitem[Roumboutsos and Kapros, 2008]{Roumboutsos2008209}
Roumboutsos, A. and Kapros, S. (2008).
\newblock A game theory approach to urban public transport integration policy.
\newblock {\em Transport Policy}, 15(4):209 -- 215.

\bibitem[Rozenblat, 2020]{rozenblat2020extending}
Rozenblat, C. (2020).
\newblock Extending the concept of city for delineating large urban regions
  (lur) for the cities of the world.
\newblock {\em Cybergeo: European Journal of Geography}.

\bibitem[Russo and Musolino, 2012]{russo2012unifying}
Russo, F. and Musolino, G. (2012).
\newblock A unifying modelling framework to simulate the spatial economic
  transport interaction process at urban and national scales.
\newblock {\em Journal of Transport Geography}, 24:189--197.

\bibitem[Sanders et~al., 2018]{sanders2018prevalence}
Sanders, J.~B., Farmer, J.~D., and Galla, T. (2018).
\newblock The prevalence of chaotic dynamics in games with many players.
\newblock {\em Scientific reports}, 8(1):1--13.

\bibitem[Schmitt, 2014]{schmitt2014modelisation}
Schmitt, C. (2014).
\newblock {\em Mod{\'e}lisation de la dynamique des syst{\`e}mes de peuplement:
  de SimpopLocal {\`a} SimpopNet.}
\newblock PhD thesis, Paris 1.

\bibitem[Scott, 2019]{scott2019city}
Scott, A.~J. (2019).
\newblock City-regions reconsidered.
\newblock {\em Environment and Planning A: Economy and Space}, 51(3):554--580.

\bibitem[See and Openshaw, 2000]{see2000hybrid}
See, L. and Openshaw, S. (2000).
\newblock A hybrid multi-model approach to river level forecasting.
\newblock {\em Hydrological Sciences Journal}, 45(4):523--536.

\bibitem[Shao et~al., 2009]{shao2009ground}
Shao, M., Zhang, Y., Zeng, L., Tang, X., Zhang, J., Zhong, L., and Wang, B.
  (2009).
\newblock Ground-level ozone in the pearl river delta and the roles of voc and
  nox in its production.
\newblock {\em Journal of Environmental Management}, 90(1):512--518.

\bibitem[Shen, 2002]{shen2002urban}
Shen, J. (2002).
\newblock Urban and regional development in post-reform china: the case of
  zhujiang delta.
\newblock {\em Progress in Planning}, 57(2):91--140.

\bibitem[Soja, 2000]{soja2000postmetropolis}
Soja, E.~W. (2000).
\newblock Postmetropolis critical studies of cities and regions.

\bibitem[Swerts, 2017]{swerts2017database}
Swerts, E. (2017).
\newblock A data base on chinese urbanization: Chinacities.
\newblock {\em Cybergeo: European Journal of Geography}.

\bibitem[Taubenb{\"o}ck et~al., 2014]{taubenbock2014new}
Taubenb{\"o}ck, H., Wiesner, M., Felbier, A., Marconcini, M., Esch, T., and
  Dech, S. (2014).
\newblock New dimensions of urban landscapes: The spatio-temporal evolution
  from a polynuclei area to a mega-region based on remote sensing data.
\newblock {\em Applied Geography}, 47:137--153.

\bibitem[Thomas et~al., 2018]{thomas2018city}
Thomas, I., Jones, J., Caruso, G., and Gerber, P. (2018).
\newblock City delineation in european applications of luti models: review and
  tests.
\newblock {\em Transport Reviews}, 38(1):6--32.

\bibitem[Timmermans, 2003]{timmermans2003saga}
Timmermans, H. (2003).
\newblock The saga of integrated land use-transport modeling: how many more
  dreams before we wake up?
\newblock In {\em Keynote paper, Moving through nets: The Physical and social
  dimension of travel, 10th International Conference on Travel Behaviour
  Research}.

\bibitem[Tretyakov et~al., 2011]{tretyakov2011fast}
Tretyakov, K., Armas-Cervantes, A., Garc{\'\i}a-Ba{\~n}uelos, L., Vilo, J., and
  Dumas, M. (2011).
\newblock Fast fully dynamic landmark-based estimation of shortest path
  distances in very large graphs.
\newblock In {\em Proceedings of the 20th ACM international conference on
  Information and knowledge management}, pages 1785--1794. ACM.

\bibitem[Verdier and Bretagnolle, 2007]{verdier2007extension}
Verdier, N. and Bretagnolle, A. (2007).
\newblock L'extension du r{\'e}seau des routes de poste en france, de 1708
  {\`a} 1833.
\newblock In {\em histoire des r{\'e}seaux postaux en Europe du XVIIIe au XXIe
  si{\`e}cle}, pages 155--193. Comit{\'e} pour l'histoire de la Poste.

\bibitem[Vincent-Geslin and Ravalet, 2016]{vincent2016determinants}
Vincent-Geslin, S. and Ravalet, E. (2016).
\newblock Determinants of extreme commuting. evidence from brussels, geneva and
  lyon.
\newblock {\em Journal of Transport Geography}, 54:240--247.

\bibitem[Waddell, 2002]{waddell2002urbansim}
Waddell, P. (2002).
\newblock Urbansim: Modeling urban development for land use, transportation,
  and environmental planning.
\newblock {\em Journal of the American planning association}, 68(3):297--314.

\bibitem[Wegener, 2004]{wegener2004overview}
Wegener, M. (2004).
\newblock Overview of land use transport models.
\newblock In {\em Handbook of transport geography and spatial systems}. Emerald
  Group Publishing Limited.

\bibitem[Wegener and F{\"u}rst, 2004]{wegener2004land}
Wegener, M. and F{\"u}rst, F. (2004).
\newblock Land-use transport interaction: state of the art.

\bibitem[Wilensky, 1999]{wilensky1999netlogo}
Wilensky, U. (1999).
\newblock Netlogo.

\bibitem[Xie and Levinson, 2009]{xie2009modeling}
Xie, F. and Levinson, D. (2009).
\newblock Modeling the growth of transportation networks: a comprehensive
  review.
\newblock {\em Networks and Spatial Economics}, 9(3):291--307.

\bibitem[Xie and Levinson, 2011a]{Xie2011}
Xie, F. and Levinson, D.~M. (2011a).
\newblock {\em Governance Choice - A Theoretical Analysis}, pages 179--198.
\newblock Springer New York, New York, NY.

\bibitem[Xie and Levinson, 2011b]{xie2011governance}
Xie, F. and Levinson, D.~M. (2011b).
\newblock Governance choice-a simulation model.
\newblock In {\em Evolving Transportation Networks}, pages 199--221. Springer.

\bibitem[Xu and Yeh, 2010a]{xu2010mega}
Xu, J. and Yeh, A. (2010a).
\newblock Mega-city region governance and planning: An international
  comparative perspective.

\bibitem[Xu and Yeh, 2005]{xu2005city}
Xu, J. and Yeh, A.~G. (2005).
\newblock City repositioning and competitiveness building in regional
  development: New development strategies in guangzhou, china.
\newblock {\em International Journal of Urban and Regional Research},
  29(2):283--308.

\bibitem[Xu and Yeh, 2008]{xu2008planning}
Xu, J. and Yeh, A.~G. (2008).
\newblock Planning the mega-city regions in china: rationales and policies.
\newblock In {\em Workshop on Governing Global City-Regions in China and the
  West: A Comparative Approach}. School of Modern Languages and Cultures, The
  University of Hong Kong and Fulbright-Hong Kong-America Center.

\bibitem[Xu and Yeh, 2010b]{xu2010governance}
Xu, J. and Yeh, A.~G. (2010b).
\newblock {\em Governance and planning of mega-city regions: An international
  comparative perspective}.
\newblock Routledge.

\bibitem[Yeh and Chen, 2020]{yeh2020cities}
Yeh, A. G.-O. and Chen, Z. (2020).
\newblock From cities to super mega city regions in china in a new wave of
  urbanisation and economic transition: Issues and challenges.
\newblock {\em Urban Studies}, 57(3):636--654.

\bibitem[Yusufzyanova and Zhang, 2011]{yusufzyanova2011forecasting}
Yusufzyanova, D. and Zhang, L. (2011).
\newblock Forecasting transportation network evolution and performance under
  existing and alternative transportation planning processes.
\newblock In {\em ICCTP 2011: Towards Sustainable Transportation Systems},
  pages 4145--4156.

\bibitem[Zhang and Levinson, 2017]{zhang2016model}
Zhang, L. and Levinson, D. (2017).
\newblock A model of the rise and fall of roads.
\newblock {\em Journal of Transport and Land Use}, 10(1):337--356.

\end{thebibliography}

%%%%%%%%%%%%%%%%%%%%%%%
\section*{Appendix A: Convergence of the Land-use submodel}

We study here the issue of the convergence in time of the distribution of activities, with a fixed infrastructure.

Let consider a very simple case: by taking $\lambda = 0$ the problem is made not spatial and by taking $\gamma_A = 1$ we achieve the decoupling between population and employments. By denoting $\beta' = \sum_j E_j \cdot \beta$ and $P_0 = \alpha \cdot \sum_i P_i$, the existence of a fixed point for populations is equivalent to solving the equation

\begin{equation}
P_i = P_0 \cdot \frac{\exp\left(\beta' \cdot P_i\right)}{\sum \exp\left(\beta' \cdot P_i\right)}
\end{equation}

The function is indeed continuous in $P_i$ and variation ranges for population are $[0,\sum_i P_i]$, it therefore admits a fixed point through the Brouwer fixed point theorem.

In all generality, if we write

\begin{equation}
(\vec{P}(t+1),\vec{E}(t+1)) = f(\vec{P}(t),\vec{E}(t))
\end{equation}

for arbitrary parameter values, the function $f$ is also continuous in each component, and takes its values with a bounded closed interval (employments being also limited) therefore a compact. The same way that \cite{leurent2014user} establish it for a model of traffic flows, we also have a fixed point in our case, what corresponds to an equilibrium point. The uniqueness is however not trivial and there is no reason for it to be a priori verified. We empirically verified the systematic convergence at fixed infrastructure (see the exploration of the parameter space in main text and extended plots on the open repository of the project).

%%%%%%%%%%%%%%%%%%
\section*{Appendix B: Probabilities to cooperate}

\subsection*{B.1 Nash Equilibrium}

The equilibrium assumption implies that conditional expectancy of each player are equal given their two choices, i.e. that

\begin{equation}
\Eb{U_i|S_i=C} = \Eb{U_i|S_i=NC}
\end{equation}

It is equivalent in that case to maximise $\Eb{U_i}$ as a function of $p_i$, since by conditioning we have$\Eb{U_i} = p_i \Eb{U_i|S_i = C} + (1 - p_i) \Eb{U_i|S_i = NC}$, and thus $\frac{\partial \Eb{U_i}}{\partial p_i} = \Eb{U_i|S_i = C} - \Eb{U_i| S_i = NC}$.

We have then

\begin{equation}
\Eb{U_i|S_i=C} = p_{1-i} U_i(S_i=C,S_{1-i}=C) + (1- p_{1-i}) U_i (S_i=C,S_{1-i}=NC)
\end{equation}

and thus

\begin{equation}
\hspace{-1cm}
\begin{split}
	p_{1-i} & U_i(S_i=C,S_{1-i}=C) + (1- p_{1-i}) U_i (S_i=C,S_{1-i}=NC) \\
	& = p_{1-i} U_i(S_i=NC,S_{1-i}=C) + (1- p_{1-i}) U_i (S_i=NC,S_{1-i}=NC)
\end{split}
\end{equation}

what gives

\begin{equation}
p_{1-i} = - \frac{U_i(C,NC) - U_i(NC,NC)}{\left(U_i(C,C) - U_i(NC,C)\right) - \left(U_i(C,NC) - U_i(NC,NC)\right)}
\end{equation}

By substituting the expressions of utilities from the payoff matrix, we obtain the expression of $p_i$ as a function of collaboration cost $J$ and of the difference of accessibility differentials.

Note that Nash equilibrium forces feasibility conditions on $J$ and accessibility gains to keep a probability. These are
\begin{itemize}
	\item $ J \leq \Delta X_{1 - i}(Z^{\star}_{C}) - \Delta X_{1 - i}(Z^{\star}_{1 - i})$, what can be interpreted as a cost-benefits condition for cooperation, e.g. that the gain induced by the common infrastructure must be larger than the collaboration cost;
	\item $\Delta X_{1 - i}(Z^{\star}_{C}) \leq \Delta X_{1 - i}(Z^{\star}_{1 - i})$, e.g. that the gain induced by the common infrastructure must be positive.
\end{itemize}

\subsection*{B.2 Discrete choice game}

%%%%%%%%%%%%%%%
\begin{figure*}
    \begin{center}
	\includegraphics[width=\linewidth]{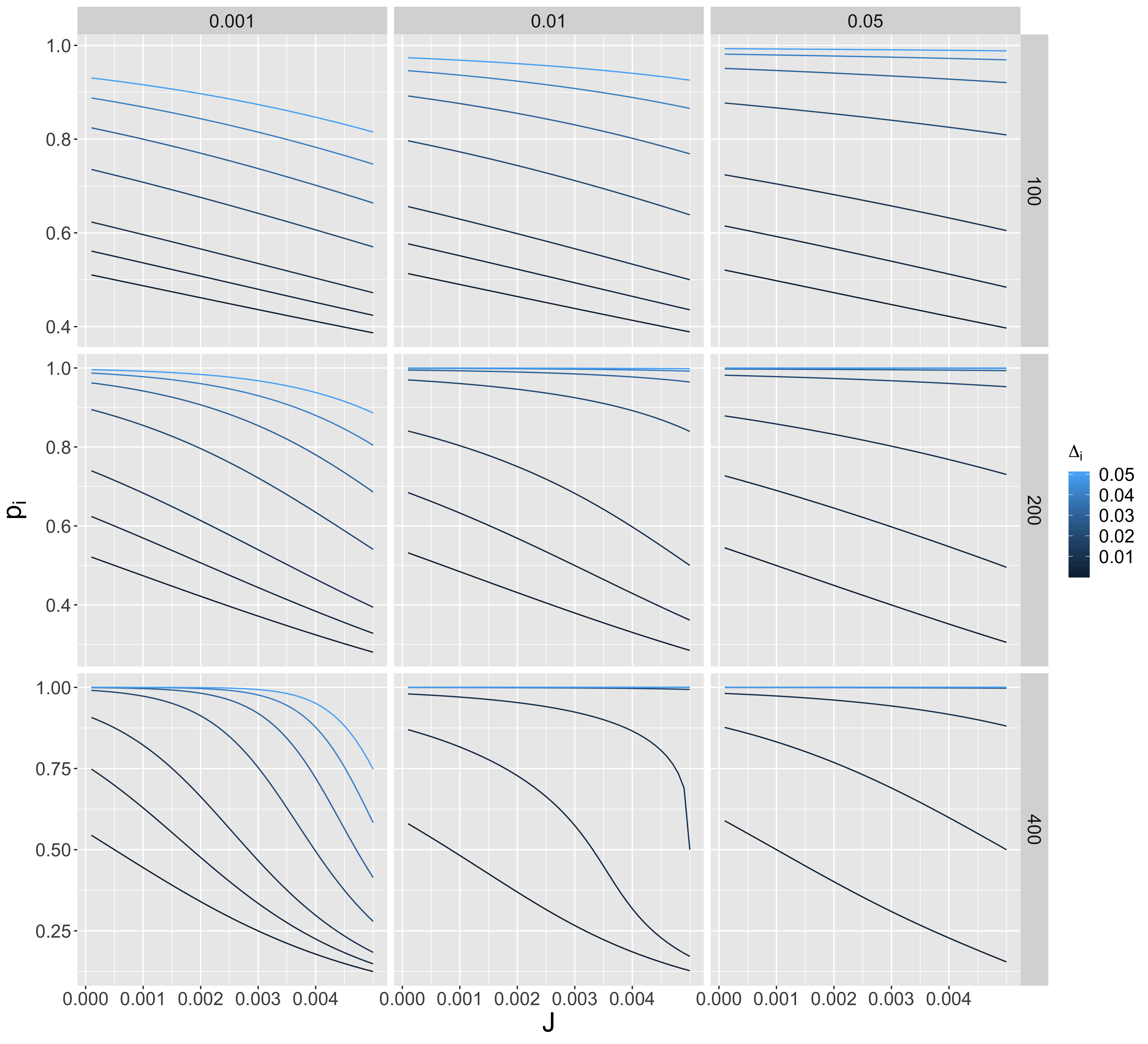}
	\end{center}
    \caption{\textbf{Empirical behavior of probabilities for the discrete choice game.} We plot numerically obtained values of $p_i$ as a function of collaboration cost $J$, for varying $\Delta_i$ (colour), $\Delta_{1-i}$ (columns) and $\beta_{DC}$ (rows).\label{fig:dcgame-empirical-probas}}
\end{figure*}
%%%%%%%%%%%%%%%

To determine the probability to cooperate in the discrete choice case, we have to solve $f(p_i) = 0$ with

\begin{equation}
\label{eq:dcproba-tosolve}
f(x) = \frac{1}{1+\exp\left[-\beta_{DC}\frac{\Delta_i}{1 + \exp(-\beta_{DC}(x \Delta_{1-i} - J))} - J\right]} - x
\end{equation}

where we wrote $\Delta_i = \Delta X_{i}(Z^{\star}_{C}) - \Delta X_{\bar{i}}(Z^{\star}_{i})$.

We have $f(0) > 0 $ and $f(1) < 0$ and the function $f$ is continuous, therefore there always exists a solution $x\in [0,1]$ by the theorem of intermediate values.

Regarding uniqueness, it can be shown under some assumptions that we obtain below. A computation of $\frac{\partial f}{\partial x}$ gives

\[
\frac{\partial f}{\partial x} = 2 (\cosh u(x) - 1) + \beta^2 \Delta_i \Delta_{1-i} \frac{\exp(-\beta_{DC}(x \Delta_{1-i} - J))}{(1 + \exp(-\beta_{DC}(x \Delta_{1-i} - J)))^2}
\]

where $u(x) = -\beta_{DC} (\frac{\Delta_i}{1 + \exp(-\beta_{DC}(x \Delta_{1-i} - J))} - J)$.

As $\cosh u \geq 1$, we have $\frac{\partial f}{\partial x} > 0$ if $\Delta_i \Delta_{1-i} > 0$. The function is in this case strictly increasing and we have a unique solution.

In practice, the solution is determined with the Brent algorithm provided by the \texttt{numanal} NetLogo extension, with boundaries $[0,1]$ and a tolerance of $0.01$.

To give an empirical insight into the behavior of probabilities for this game, we solve numerically Eq.~\ref{eq:dcproba-tosolve} for different values of parameters $\Delta_i, \Delta_{1-i}, J, \beta_{DC}$. These solutions are shown in Fig.~\ref{fig:dcgame-empirical-probas}. We find an always decreasing behavior as a function of collaboration cost $J$. Large values of $\beta_{DC}$ imply a stronger non-linear behavior. Lower values of $\Delta_i$ significantly decrease the probability $p_i$, while $\Delta_{1-i}$ has a more moderate influence.

\section*{Appendix C: Implementation}

\subsection*{C.1 Distance matrix}

The distance matrix is updated in a dynamical way because of execution time issues (given the number of network updates), through the following procedure:
\begin{enumerate}
	\item The euclidean distance matrix $d(i,j)$ is computed analytically
	\item The shortest paths between intersections of links (between cells of the corresponding raster network) are updated in a dynamical way (step of complexity $O(N_{inters}^3$):
	\begin{itemize}
		\item For each new intersection, shortest paths towards all other intersections are computed using the old matrix and the new link.
		\item For all former shortest paths, they are updated if needed after checking potential shortcuts using the new link.
		\item The correspondence between basic network cells and intersections is updated.
	\end{itemize}
	\item Connected components and the distances between them are updated (complexity in $O(N_{nw}^2)$)
	\item Distances within the network between network cells are updated, with the heuristic of minimal connections only (a unique shortest link between each cluster) (complexity in $O(N_{nw}^2)$)
	\item Effective distances between all cells (taking speed into account and congestion if it is implemented) are computed as the minimum between euclidean distance and \[\min_{C,C'}{d(i,C)+d_{nw}(p_C(i),p_C'(j))+d(C',j)}\] that we approximate by taking $\min_C$ only in the implementation, what is consistent with relatively interaction ranges that we consider. The complexity is in $O(N_{clusters}^2\cdot N^2)$.
\end{enumerate}

\subsection*{C.2 Network growth}

The potential infrastructures, with a count of $N_I$ during the heuristic search of an optimal infrastructure, are randomly drawn among all possible infrastructures having an extremity at the centre of a cell. If the extremity is at a distance lower than a threshold of an already existing link of the network, it is replaced by its projection on the corresponding link. This is a snapping step which allows to obtain a network with a reasonable form locally. In correspondence with the raster representation of the network, we take $\theta_I = 1$, what corresponds to the size of a cell.

\end{document}